\documentclass[twoside]{article}

\usepackage{float}
\usepackage{color}
\usepackage{greekbf}
\usepackage{float}
\usepackage{amsmath,amssymb}
\usepackage{mathrsfs}
\usepackage{enumerate}
\usepackage{srcltx}
\usepackage{mathtools}
    \mathtoolsset{showonlyrefs}
\usepackage{graphicx, psfrag}
\usepackage{hyperref}
\hypersetup{bookmarks=false,
colorlinks=false,
linkcolor=black,
urlcolor=black,
pagebackref=true,
pdfstartview={FitH},
bookmarksopenlevel=2
}

\newtheorem{remark}{Remark}

\DeclareMathAlphabet{\mathbf}{OT1}{cmr}{bx}{it}

\definecolor{red}{rgb}{0.9,0,0}
\definecolor{blue}{rgb}{0.2,0.2,0.8}
\definecolor{green}{rgb}{0.0,0.5,0.2}
\definecolor{darkblue}{rgb}{0.2,0.2,0.5}
\definecolor{orange}{rgb}{1,0.5,0}

\newcommand {\csib}{\mathbf{\xi}}
\newcommand {\xib}{\mathbf{\xi}}
\newcommand {\fb} {\mathbf{f}}
\newcommand {\qb} {\mathbf{q}}
\newcommand {\rb} {\mathbf{r}}
\newcommand {\chib}{\mathbf{\chi}}
\newcommand {\Bb} {\mathbf{B}}
\newcommand {\Ab} {\mathbf{A}}
\newcommand {\nb} {\mathbf{n}}
\newcommand {\hb} {\mathbf{h}}
\newcommand {\ab} {\mathbf{a}}
\newcommand {\0} {\textbf{0}}
\newcommand {\Tc}  {\mathcal{T}}
\newcommand {\cb} {\mathbf{c}}
\newcommand {\db} {\mathbf{d}}
\newcommand {\bb} {\mathbf{b}}

\usepackage{arpione}

\begin{document}

\title{\vspace{-3cm} {\bf Geometry and Self-Stress of Single-Wall Carbon Nanotubes and Graphene via a Discrete Model Based on a 2nd-Generation REBO Potential}}

\author{
Antonino Favata$^1$\!\!\!\!\! \and Andrea Micheletti$^2$\!\!\!\!\! \and Paolo Podio-Guidugli$^{3,4}$\!\!\!\!\! \and  Nicola M. Pugno$^{1,5,6}$
}

\date{\today}

\maketitle

\vspace{-1cm}
\begin{center}
{\small
$^1$ Laboratory of Bioinspired and Graphene Nanomechanics\\ Department of Civil, Environmental and Mechanical Engineering\\ University of Trento, Italy\\[2pt]
\href{mailto:antonino.favata@unitn.it}{antonino.favata@unitn.it}, \href{mailto:nicola.pugno@unitn.it}{nicola.pugno@unitn.it}\\[8pt]
$^2$ Dipartimento di Ingegneria Civile e Ingegneria Informatica\\
University of Rome TorVergata, Italy\\[2pt]
\href{mailto:micheletti@ing.uniroma2.it}{micheletti@ing.uniroma2.it}\\[8pt]
$^3$ Accademia Nazionale dei Lincei\\
Rome, Italy\\[5pt]
$^4$ Department of Mathematics\\University of Rome TorVergata, Italy \\[2pt] 
\href{mailto:ppg@uniroma2.it}{ppg@uniroma2.it}\\[8pt]
$^5$ Center for Materials and Microsystems\\
Fondazione Bruno Kessler,  Trento, Italy\\[8pt]
$^6$ School of Engineering and Materials Science\\
Queen Mary University of London,  UK
}

\end{center}

\pagestyle{myheadings}
\markboth{A.~Favata, A.~Micheletti, P.~Podio-Guidugli, N.M.~Pugno}
{Geometry and Self-Stress of Single-Walled CNTs}

\vspace{-0.5cm}
\section*{Abstract}

The main purpose of this paper is to evaluate the \emph{self-stress} state of single-wall carbon nanotubes (CNTs)  and flat graphene strips (FGSs) in their \emph{natural} equilibrium state, that is, the state prior to the application of external loads. We model CNTs as discrete elastic structures, whose shape and volume changes are governed by a Reactive Empirical Bond-Order (REBO) interatomic potential of second generation. The kinematical  variables we consider are bond lengths, bond angles, and dihedral angles; to changes of each of these variables we associate a work-conjugate \emph{nanostress}. To determine the self-stress state in a given CNT, we formulate the load-free equilibrium problem as a minimum problem for the interatomic potential, whose solution yields the equilibrium nanostresses; next, by exploiting the  nonlinear constitutive dependence we derive for nanostresses in terms of a list of kinematical variables, we determine the equilibrium values of the latter; finally, from the equilibrium values of the kinematical variables we deduce the natural geometry and, in particular, the \emph{natural radius}. 

Our theoretical framework accommodates CNTs of whatever chirality. In the achiral case, when we can count on maximal intrinsic symmetries and hence the number of independent unknowns is reduced to a minimum, the stationarity conditions implied by energy minimization are relatively easy to derive and solve numerically; for chiral CNTs, we prefer to solve the minimum problem directly.

The natural-radius predictions we achieve  within our discrete-mechanics framework are in 
good agreement with the results of DFT- and TB-based calculations; the same is true for our predictions of the \emph{self-energy}, that is, the energy associated with self-stress (called \emph{cohesive energy} in the literature); we surmise that our discrete mechanical model may serve as a source of benchmarks for MD simulation algorithms.

We find that self-stress depends on changes in both bond and dihedral angles in achiral CNTs and, in addition,  on changes in bond length in chiral CNTs. Our analysis applies also to FGSs, whose self-stress and self-energy we  evaluate; we find that, in FGSs self-stress is associated exclusively with changes in bond angle. 

\section*{Keywords}
Carbon Nanotubes; Graphene; Discrete Model; REBO Potential; Dihedral Angles; Nanostress; Self-Stress; Benchmark

\tableofcontents
\newpage

\section{Introduction}

In this paper we propose a discrete mechanical model to investigate whether carbon nanotubes (CNTs)  and  flat graphene strips (FGSs) are in a state of \textit{self-stress} in their \emph{natural} configuration, that is, in a state of  mechanical coaction prior to  application of external loads. That all pristine CNTs are somehow and to some extent  self-stressed is implied by the collection of unzipping experiments that have been performed with various techniques ever since the pioneer study \cite{Kosynkin_2009} was published in 2009.\footnote{While the possibility of obtaining graphene ribbons from
CNTs is experimentally proved, there is not yet a conclusive evidence that completely flat graphene sheets can be obtained by unzipping procedures (see e.g. \cite{Wang2012}).}
We not only show that all CNTs, whatever their size and chirality, suffer a self-stress, but we also
identify its different sources, as many as the types of kinematical variables we consider. In addition to changes in bond lengths and bond angles, our energy form accounts for changes in dihedral angles; to each of these kinematical variables we associate a work-conjugated \textit{nanostress}.\footnote{Changes in dihedral angles have been  ignored so far in most of discrete models of carbon allotropes, a noticeable exception being the series of studies \cite{Georgantzinos_2010,Georgantzinos_2011,Giannopoulos_2011,Giannopoulos_2012}.} Our quantitative evaluation of nanostresses in CNTs is achieved with the use of Brenner's 2nd-generation Reactive Empirical Bond-Order (REBO) interatomic potential \cite{Brenner_2002}; by taking the infinite-radius limit we also determine the self-stress state of FGSs. Both for CNTs and FGSs, we find that the self-stress is of the same order of magnitude as the stress induced by traction loads that are about two thirds of the corresponding fracture loads in the case of FGSs and that, in the scarce light of the information we gather from the few experiments available, are comparable with fracture loads in the case of CNTs. This fact strongly suggests that self-stress be properly incorporated in whatever model of carbon allotropes, be it discrete or continuum.

Discrete  models have been used since long to predict the mechanical behavior of CNTs and graphene. The linear model exploited in \cite{Geng2006} to obtain closed-form expressions for the elastic properties of armchair (A) and zigzag (Z) CNTs has been extended in \cite{Xiao2005} to study torsion loading, with  nonlinearities handled by means of a modified Morse potential.  A similar approach has been used in \cite{Shen_2004} to investigate various loading conditions, and in \cite{Wang2004} to evaluate  effective in-plane stiffness and bending rigidity of A- and Z-CNTs. In \cite{Chang_2005}, the model of \cite{Chang_2003} is extended to chiral  CNTs, an issue addressed also in \cite{Chang_2006}. 
Recently,   a discrete model allowing to treat general load conditions, arbitrary chirality, and an initially stressed state, has been proposed in \cite{Merli2013}, and a  geometrically nonlinear theory of  discrete elastic structures accounting for the effects of self-stress has been presented in \cite{Favata_2014}. Discrete mechanics has also  served as a scale-bridging tool to build shell theories \cite{Chang_2010,Bajaj_2013,Favata2014a}. On the computational side,  the  methods presented in \cite{Meo2006} are worth-mentioning, where nonlinear torsional spring elements are adopted and implemented in a FE code. The mechanical properties of graphene sheets and ribbons have been analyzed with a similar approach; in particular, FE formulations employing both linear \cite{Georgantzinos_2010} and nonlinear springs  have been given \cite{Geng2006,Georgantzinos_2011,Giannopoulos_2011,Giannopoulos_2012}.

The size of a CNT of given chirality is customarily specified in terms of its \emph{nominal} radius, that is, the radius that it would have if it were possible to manufacture it by rolling a monolayer FGS up with no energy expenditure; this so-called Rolled-Up Model (RUM) was proposed in the nineties just after CNTs came to the fore and, curiously enough, way before graphene did.
 The nominal radius is easy to compute; it approximates the natural radius from below, poorly for big curvatures, better and better for larger and larger CNTs.
 Now, to determine the self-stress state in a given CNT, we formulate the load-free equilibrium problem as a minimum problem, whose solution yields the equilibrium nanostresses. The minimum of the interatomic potential is not realized in the nominal configuration; instead, it is realized in the natural configuration.  In a sense, the natural geometry of a CNT is an intermediate product of our procedure to determine the self-stress state.

 On having recourse to RUM, a CNT's circumference and length are taken equal to, respectively, the width and the length of the `parent' FGS. One intrinsic approximation of RUM is that bond lengths come out shorter in CNTs than in their `parent' FGSs, due to the difference between the length of a helix segment and the distance between its endpoints \cite{Kurti2003}; consequently, RUM is accurate only for CNTs of rather large radius (about three times the length of a C bond in graphene).
Several studies have been performed to determine in a more precise way the geometry of CNTs of small radius. Some studies adopt a DFT  approach \cite{Kanamitsu2002,Machon2002,Kurti2003,Cabria2003,Budyka_2005,Demichelis_2011}, others a TB \cite{Popov2004} or an interatomic potential approach \cite{Jiang_2003,Jindal_2008}; in all cases, the `relaxed' ($\equiv$ natural) configuration of a CNT is determined through energy minimization, starting from the `unrelaxed' ($\equiv$ nominal) configuration furnished by  RUM.
The main findings of these studies can be summarized as follows: whatever the chirality, (a) `relaxed' and `unrelaxed' bond lengths and bond angles are different; (b) `relaxed' CNTs have larger radius. Interestingly,  in \cite{Lee2011} geometrical relationships alternative to those of RUM are derived, in order to obtain the radius of a CNT once its bond lengths and bond angles are known.

Our theoretical framework accommodates CNTs of whatever chirality. We show that the symmetries intrinsic to \emph{achiral} CNTs induce a drastic reduction of the number of kinematic unknowns needed in case of arbitrary chirality. This greatly simplifies the formal developments leading to achieve energy minimization \emph{via} the solution of a set of \emph{stationarity conditions in terms of equilibrium nanostresses}, because for achiral CNTs those conditions are few and relatively easy to obtain; moreover, once the nonlinear dependence of nanostresses on the kinematic unknowns is taken into account, numerical computations turn out to be reasonably light. As to \emph{chiral} CNTs, we prefer to perform energy minimization directly. We derive the explicit form of constitutive equations for nanostresses in this case as well, and we provide a body of quantitative results showing that \emph{self-energy} (that is, the energy associated with self-stress, called \emph{cohesive energy} in the literature) and self-stress of chiral CNTs are close to those of achiral CNTs of similar radius (see Sect. \ref{chiral}).
We find that self-stress depends on changes in both bond and dihedral angles in achiral CNTs and, in addition,  on changes in bond length in chiral CNTs. Our analysis applies also to FGSs, whose self-stress and self-energy
we  evaluate; we find that, in FGSs self-stress is associated exclusively with changes in bond angle.
The natural-radius predictions we achieve  within our discrete-mechanics framework are in
good agreement with the results of DFT- and TB-based calculations; the same is true for our predictions of the \emph{self-energy}.

 A limitation of our study is that our findings refer to \emph{defectless} CNTs and FGSs (for an assessment of how defects affect elastic response and strength of CNTs, see \cite{Pugno_2007, Pugno_2007_a}).
An element of  novelty is that we have chosen
an interatomic potential that, to the best of our knowledge, has not been employed before in connection with CNTs of small radius; the potential we use  features a contribution due to changes in dihedral angles, so as to account for the curvature-related effects of electrons' orbital distortion and rehibridization. For this reason, we believe that our predictions are potentially better -- in the sense that they can turn out to be closer to the results obtained by  DFT or TB approaches -- than to those obtained when first-generation potentials  \cite{Tersoff_1988,Tersoff_1989,Brenner_1990} are employed. In fact, we envisage the possibility of tuning the parameters from which our potential  depends so as to minimize the discrepancies between our  predictions and DFT's or TB's, with a view towards improving the performances of MD simulations based on such a finely tuned potential. And, even in the absence of such tuning, we believe that our discrete
mechanical model, which leads to a quite standard energy minimization procedure requiring negligible computational  time, may serve as a source of benchmarks for MD simulation algorithms:  as is, in case those algorithms incorporate the same intermolecular potential; after modest adjustments, in case of other REBO potentials.

In our opinion, recognizing that both CNTs and FGSs are self-stressed structures and, morever,  qualifying and quantifying their self-stress state is the
main element of novelty of our study. Accounting for self-stress is crucial to produce reliable predictions whenever a harmonic approximation of the interatomic potential is accepted \cite{Favata_2014}; see also \cite{Xu2013}, where it is shown that bond-angle self-stress can contribute to the bending stiffness of monolayer FGSs.
We find that the dihedral contribution to self-energy is large, about half of the total for CNTs of large radius, a result consistent with those in \cite{Lu_2009}; we also find that the dihedral contribution is less important in small-radius CNTs, a fact that can be justified by the large change of bond angles when curvature is large.

We consider a generic molecular aggregate, kept together by a system of conservative intermolecular forces; we describe such an aggregate as a discrete mechanical structure, whose configuration is identified by a finite list of order parameters; and we determine the conditions of \textit{natural equilibrium} for such an aggregate. In Section \ref{geokin}, we focus on hexagonal carbon lattices: we detail their geometry and kinematics and we recapitulate the \textit{nominal geometry} of achiral CNTs, that is, their geometry as viewed according to the RUM.  We then deal with the \textit{exact geometry}, after having chosen the proper order parameter string,  in the light of the developments of Section \ref{natequi}.
In Section \ref{Equil}, we derive the equations governing the natural equilibrium of achiral CNTs, while, for chiral CNTs, we give the form of the energy to be minimized, providing in both cases the constitutive relations for the nanostresses. In the final Section \ref{discuss}, we present and discuss the results of our theory and make a comparison with the literature.   Four Appendices complete the paper: the first contains certain geometrical and analytical details about the computation of dihedral angles that, although indispensable, would have made unduly heavier the relative developments in the main text; the second contains a reasoned presentation of the 2nd-generation Brenner potential, in its general form and in the version we use; the third  contains the balance equations for the case of uniform traction of CNTs and FGSs, a case we use to compare energies and stresses induced by an external load; the fourth contains tables collecting the data used to draw the plots in Section \ref{discuss}.

\section{Equilibria of discrete structures}\label{natequi}
The discrete mechanical structures we study consist of interacting C atoms occupying a finite hexagonal lattice; their configurations are determined by assigning admissible positions in space to all lattice points; each configuration has an energetic cost, computable by evaluating a given energy functional, which  depends in principle on the distances of all pairs of lattice points. The REBO potentials developed by Tersoff \cite{Tersoff_1988,Tersoff_1989} and Brenner \cite{Brenner_1990} have  been widely used in MD simulations of carbon-based materials; they accommodate multibody interactions up to second nearest neighbors.  A so-called \emph{2nd-generation Brenner potential} \cite{Brenner_2002} is a REBO potential that accommodates  third-nearest-neighbor interactions as well, through a bond order associated with  dihedral angles; we delineate its analytic  features in Appendix B.

In this paper we exploit the simplifications intrinsic to the highly symmetric shapes of CNTs and FGSs so as to give the 2nd-generation Brenner potential we use a form that depends  on a finite list of  
\emph{order parameters}. The latter are defined with reference to an aggregate of two or more \emph{adjacent} lattice points; when all admissible aggregates are considered, we end up with an \emph{order-parameter string} $\xib$; we call \emph{natural equilibria}  the local minima of an energy functional 
depending on $\xib$.

For low-symmetry structures, the order-parameter string one needs can be very long; this is not the case for CNTs and FGS,  especially so when natural equilibria are sought: in fact, in Section \ref{geokin}, we show that a 3-entry string $\qb$ of Lagrangian coordinates is enough to determine $\xib$.
All in all,
\begin{equation}\label{Vpot}
V=\widehat V(\qb),\quad\textrm{with}\quad \widehat V=\widetilde V\!\circ\widehat\xib\,,\quad \xib=\widehat\xib(\qb)\,.
\end{equation}

The functional $\widehat V$ we use will be introduced in Section \ref{Equil}. In absence of external forces, natural equilibria are the stationary points of $\widehat V$; any such point  $\qb_0$ satisfies
\begin{equation}\label{staz}
\delta V=\partial_\qb \widehat V(\qb_0)\cdot\delta\qb=0\quad \textrm{for all variations}\; \,\delta\qb=\qb-\qb_0,
\end{equation}
with
\[
\partial_\qb \widehat V(\qb_0)\cdot\delta\qb=\big(\partial_\qb \widehat\xib(\qb_0)\big)^T \partial_\xib \widetilde V(\xib_0)\cdot\delta\qb
=\partial_\xib \widetilde V(\xib_0)\cdot \big(\partial_\qb \widehat\xib(\qb_0)\big)\delta\qb\,,\quad \xib_0=\widehat\xib(\qb_0)\,;
\]
an equilibrium $\qb_0$ is \emph{stable} if the Hessian $\partial^2_\qb \widehat V$ is positive definite at $\qb_0$.

We set $\,\widetilde\chib:=\partial_\xib \widetilde V$, and call $\,\chib=\widetilde\chib(\xib)$ the \emph{stress mapping}, in that,
for $\,\delta\xib:=\xib-\xib_0$ the \emph{strain increment} in passing from the configuration $\xib_0$ to the configuration $\xib$,
\begin{equation}\label{increV}
\delta V=\chib\cdot\delta\xib
\end{equation}
can be regarded as the \emph{incremental expenditure of internal power}. We also set $\,\widehat\Bb:=\partial_\qb\widehat\xib$,
and  call $\widehat\Bb$ the \emph{kinematic compatibility operator}, in that
\[
\delta\xib=\widehat\Bb(\qb)\delta\qb\,.
\]
Finally, we call $\widehat\Ab:=\widehat\Bb^T$ the \emph{equilibrium operator}, and note that \eqref{staz} holds if and only if
\begin{equation}\label{ecco}
\widehat\Ab(\qb_0)\widetilde\chib(\xib_0)=\0\,,\quad \xib_0=\widehat\xib(\qb_0)\,;
\end{equation}
provided $H_0:=\partial^2_\qb \widehat V(\qb_0)$ is positive definite, $V_0:=\widehat V(\qb_0)$ is the \emph{natural binding energy}, and $\chib_0:=\partial_\xib \widetilde V(\widehat\xib(\qb_0))$ the \emph{self-stress} at the natural equilibrium  $\qb_0$.

In the presence of external loads, the functional $\widehat{V}$ has to be replaced with
\begin{equation}\label{totpot}
E=\widehat{E}(\qb):=\widehat{V}(\qb)-\widehat{P}(\qb), \quad \widehat{P}(\qb):=\fb\cdot\widehat{\db}(\qb),
\end{equation}
where the \textit{dead load} $\fb$ is work-conjugate to the generalized displacement $\widehat{\db}(\qb)$; an equilibrium point $\qb_0$ satisfies the condition:
\begin{equation}
\delta E=\partial_\xib \widetilde V(\xib_0)\cdot \big(\partial_\qb \widehat\xib(\qb_0)\big)\delta\qb-\big(\partial_\qb\widehat{\db}(\qb_0) \big)^T\fb\cdot\delta\qb=0,
\end{equation}
whence the balance equation
\begin{equation}\label{balc}
\widehat\Ab(\qb_0)\widetilde\chib(\xib_0)=\big(\partial_\qb\widehat{\db}(\qb_0) \big)^T\fb;
\end{equation}
in Appendix C, we  have recourse to this general equation to solve the traction problem.

\section{Geometry and kinematics of hexagonal carbon lattices}\label{geokin}
When regarded as discrete mechanical structures, FGSs and single-walled CNTs can be modeled in one and the same manner, because they all are  carbon allotropes with hexagonal lattices. In fact, while hereafter we focus on CNTs, our description of their geometry and kinematics applies with minimal changes to FGSs.
\subsection{Bond-related kinematic variables}
In this section, we introduce the kinematic variables associated with the interatomic bonds involving first, second and third nearest neighbors of any given atom.

\begin{figure}[h]
\centering
\includegraphics[scale=1]{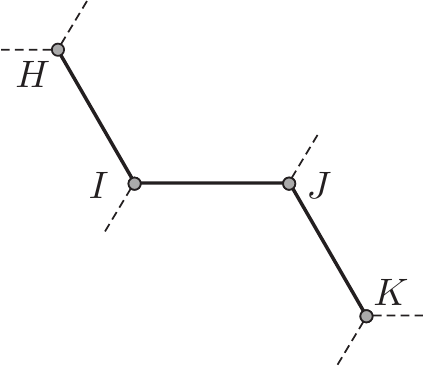}
\caption{Four C atoms of an hexagonal lattice, not necessarily planar.}
\label{4toms}
\end{figure}

With the help of Figure \ref{4toms},
consider the bond chain going from atom $H$ to atom $K$. In this chain, atoms $H$ and $J$ are the first nearest neighbors, and $K$ the second nearest neighbor, of atom $I$; moreover, atom $K$ is the third nearest neighbor of atom $H$. On denoting by $\rb_I$ the position vector of atom $I$ with respect to a chosen origin, let
\begin{equation}\label{bl}
\rb_{IJ}:=\frac{1}{r_{IJ}}\, (\rb_J-\rb_I), \qquad r_{IJ}:=|\rb_J-\rb_I|;
\end{equation}
here, $\rb_{IJ}$ is the $IJ$-bond vector, $r_{IJ}$ is the $IJ$-\emph{bond length}, the length of the covalent bond between atoms $I$ and $J$. Two bond vectors  $\rb_{HI}$ and $\rb_{IJ}$ span a plane, whose normal is:
\begin{equation}
\nb_{HIJ}:=\frac{\rb_{HI}\times\rb_{IJ}}{|\rb_{HI}\times\rb_{IJ}|};
 \end{equation}
their angle is:
\begin{equation}\label{teta}
\theta_{HIJ}:=\arccos(\rb_{HI}\times\rb_{IJ}),
\end{equation}
the $IJK$-\emph{bond angle}. Finally, the $HIJK$-\emph{dihedral angle}
\begin{equation}\label{Theta}
 \Theta_{HIJK}:=\arccos(\nb_{HIJ}\cdot\nb_{IJK})
 \end{equation}
is the angle between the planes spanned by the two pairs of bond vectors $\rb_{IJ}$, $\rb_{JK}$ and $\rb_{JI}$, $\rb_{IH}$.

 Bond length, bond angles, and dihedral angles, are the kinematic variables we consider; their dependence on the positional coordinates of the related aggregates of structure points is specified by definitions \eqref{bl}-\eqref{Theta}; we reiterate that all their changes have an energetic cost to be computed when a form for the mapping $\widetilde V$ is specified.
\subsection{The nominal geometry of achiral CNTs}\label{nom}
\begin{figure}[h]
\centering
\includegraphics[scale=0.9]{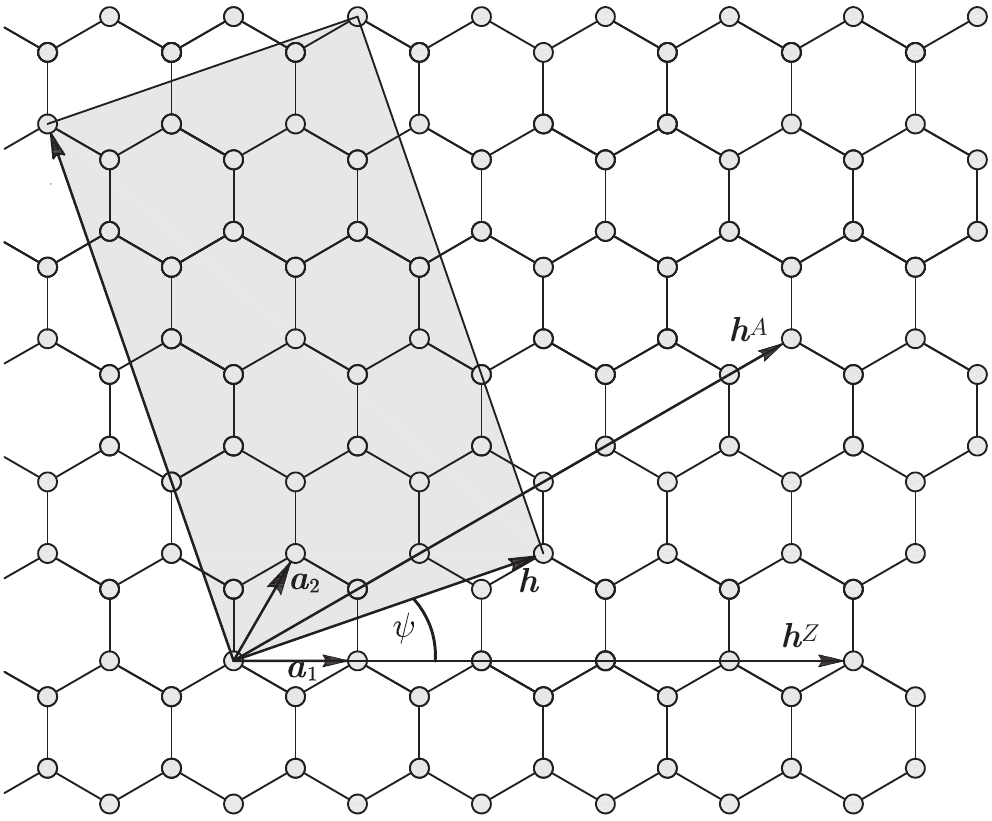}
\caption{The graphene part involved in rolling up a  chiral CNT.}
\label{1chiral}
\end{figure}

In imagination, a single-walled CNT can
be obtained by rolling and zipping up into a cylindrical shape a strip of
graphene,  that is, of a  monolayer carbon allotrope with the atomic structure of a two-dimensional flat Bravais lattice with hexagonal unit cell.  There are infinitely many ways to roll a
graphene strip up, sorted by introducing a geometrical object, the
\emph{chiral vector}:
\begin{equation}\label{chiv}
\hb=n\ab_1+m\ab_2,\quad n\geq m,
\end{equation}
where $n,m$ are integers, and $\ab_1, \ab_2$ are  \emph{lattice vectors}, such as those at a mutual angle of $\pi/3$ radians shown in Fig. \ref{1chiral};
the chiral vector forms with $\ab_1$ the \emph{chiral angle} $\psi=\arctan\big(\sqrt3\, m/(2n+m)\big)$.
When $n>m>0$, the CNT in question is termed \emph{chiral}. The  \emph{nominal radius} $\rho_0$ of a $(n,m)$-CNT is defined to be the radius of the cylinder on which the centers of the C atoms would be placed after an ideal rolling-up operation entailing no energy expenditure for the inevitable distortion of the C-C bonds; according to this definition,
\begin{equation}\label{geonec}
\rho_0=\widehat\rho_0(n,m):=\frac{\sqrt 3}{2\pi}\,n\sqrt{1+m/n+(m/n)^2}\,r_0,
\end{equation}
where $r_0$ is the length of the graphene C-C bond.

There are two types of  \emph{achiral} CNTs, namely,  $(n,0)$-\emph{zigzag} and
$(n,n)$-\emph{armchair} CNTs; in Fig. \ref{1chiral}, their chiral vectors are denoted by, respectively, $\hb^{Z}\equiv\ab_1$ ($\psi^{Z}=0$ radians) and $\hb^A$ ($\psi^{A}=\pi/6$ radians); their nominal radii are, respectively, $\rho_0^Z(n)=(\sqrt 3/2\pi)n\,r_0$ and $\rho_0^A(n)=(3/2\pi)n\,r_0$, so that
\begin{equation}\label{ineq}
\rho_0^Z(n)<\widehat\rho_0(n,m)<\rho_0^A(n),\quad n>m>0.
\end{equation}

Visualization of the rolling-up procedure is especially easy in the case of achiral CNTs; the double inequality \eqref{ineq} gives us some confidence that qualitative predictions about the natural geometry  of $(n,m)$-chiral CNTs could be made on the basis of the corresponding results for $(n,0)$-zigzag and $(n,n)$-armchair CNTs. In Section \ref{chiral}, we shall discuss how the self-stress state of a set of chiral and achiral CNTs depend on curvature.
\subsection{FGSs as  unzipped and unrolled achiral CNTs}
\begin{figure}[h!]
\centering
\includegraphics[width=1\textwidth]{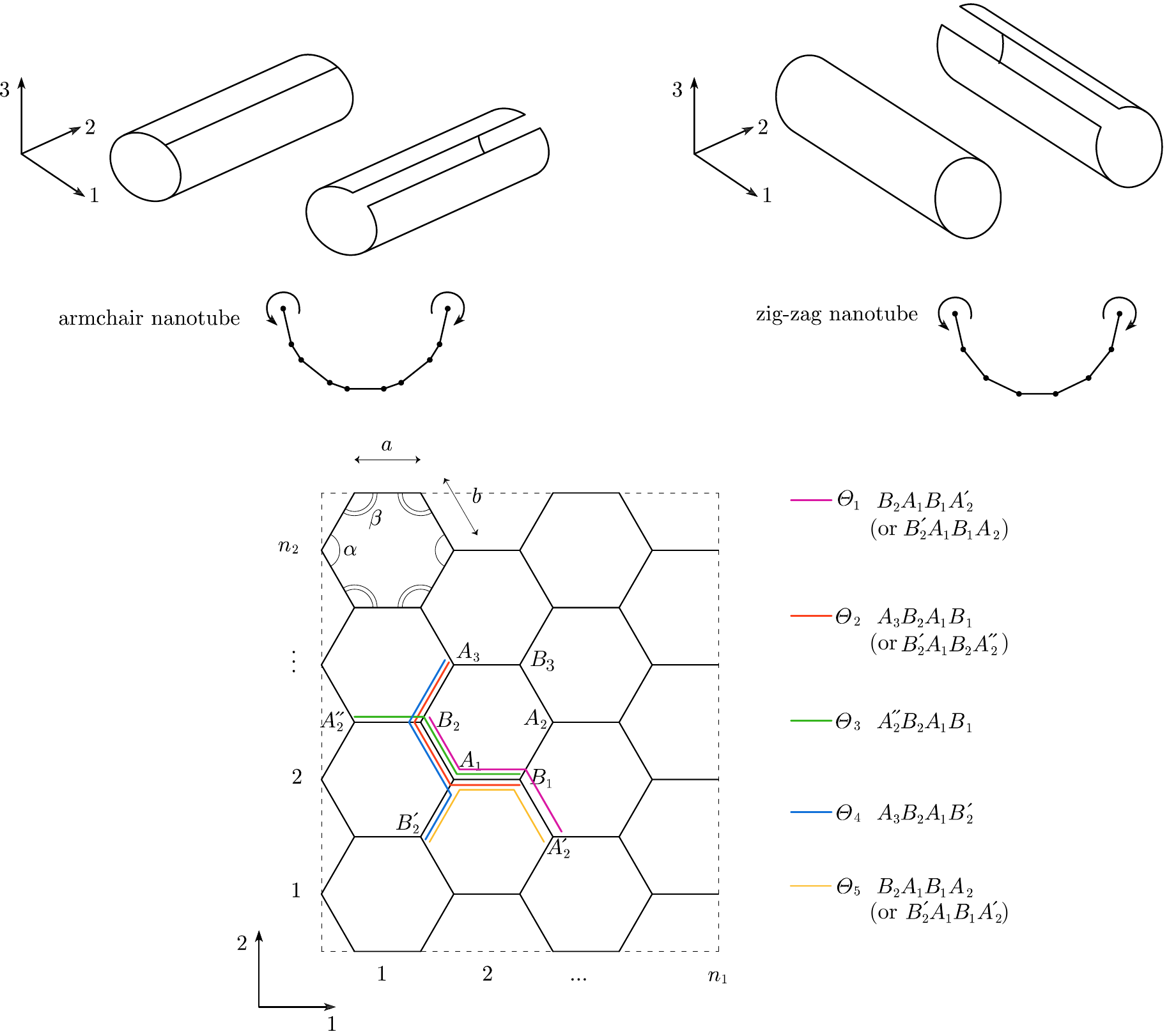}
\caption{Achiral CNTs, ideally unzipped and unrolled.}
\label{rectangle}
\end{figure}

The FGS 
depicted in the bottom part of Fig.~\ref{rectangle} 
%
is thought of as obtained by unzipping and unrolling, in imagination, the A- and a Z-CNT depicted on top, whose  axes are parallel, respectively, to the directions of axes $2$ and $1$, so as to have their chiral vectors $\hb^A$ and $\hb^Z$ aligned with directions $1$ and $2$.
Our considerations to follow hinge on  well-known intrinsic symmetries of FGSs and achiral CNTs.
The FGSs we consider consist of $n_1$ hexagonal cells in direction $1$ and of $n_2$ hexagonal cells in direction $2$; for $(n,n)$-CNTs, we set $n_1=2n$, and assume that $n_2>>n_1$; for $(n,0)$-CNTs, we set $n_2=n$ and $n_1>>n_2$. 

On looking at the representative cell $A_1 B_1 A_2 B_3 A_3 B_2 A_1$, we see that the sides $\overline{A_1 B_1}$ and $\overline{A_3 B_3}$ are aligned with $\hb^A$; we denote their common length by $a$, and call them {\em a-type bonds}; we also see that the other four sides have equal length $b$ ({\em b-type bonds}; see the cell located at the upper left corner of the strip).
 As to bond angles, they can be of \emph{$\alpha$-type}  and \emph{$\beta$-type}  (e.g., respectively, $\widehat{A_3  B_2 A_1}$ and $\widehat{B_2 A_1 B_1}$; see the upper left cell again).
 There are only five types of dihedral angles $(\Theta_1,\ldots,\Theta_5)$, which can be individuated with the help  of  the colored bond chains.
 In conclusion, the information carried by the 9-entry \emph{order-parameter substring}
 \begin{equation}\label{subs}
 \xib_{sub}:=(a,b,\alpha,\beta,\Theta_1,\ldots,\Theta_5)
 \end{equation}
 is enough to determine the deformed configuration of a representative hexagonal cell, no matter if that cell belongs to a FGS or to an achiral CNT.

Case-specific order-parameter strings might be obtained by exhaustive sequential juxtaposition, without information redundancies, of appropriately chosen substrings. However, as discussed later in Section \ref{Equil}, the total binding energy of a CNT in a natural equilibrium configuration can be found by no-redundancy summation over diatomic bonds of their individual contributions, each of which depends also on the presence of certain related bond and dihedral angles. In anticipation, we record here which and how many of these angle  variables are related to the one and the other type of diatomic bonds according to the 2nd-generation REBO potential we are going to use:
\vskip 2pt
\noindent each $a$-type bond is related to four $\beta$-type bond angles and two dihedral angles (in the case of $\overline{A_1 B_1}$, the bond angles in question are $\widehat{B_2 A_1 B_1}$, $\widehat{B_2^\prime A_1 B_1}$, $\widehat{A_1 B_1 A_2}$, and $\widehat{A_1 B_1 A_2^\prime}$; the dihedral angles are $\Theta_1$ and $\Theta_5$); each $b$-type bond  is related to two $\alpha$-type and two $\beta$-type bond angles, and to three dihedral angles (in the case of $\overline{B_2 A_1}$, the $\alpha$-type bond angles are $\widehat{A_3 B_2 A_1}$ and $\widehat{B_2 A_1 B_2^\prime}$; the $\beta$-type bond angles are $\widehat{A_2^{\prime\prime}B_2 A_1}$ and $\widehat{B_2 A_1 B_1}$; and the dihedral angles are $\Theta_2$, $\Theta_3$, and $\Theta_4$).

%
%
\subsection{Order-parameter substrings}\label{OPS}
%
%
\subsubsection{A-CNTs}
\begin{figure}[h]
\centering
\includegraphics[scale=0.95]{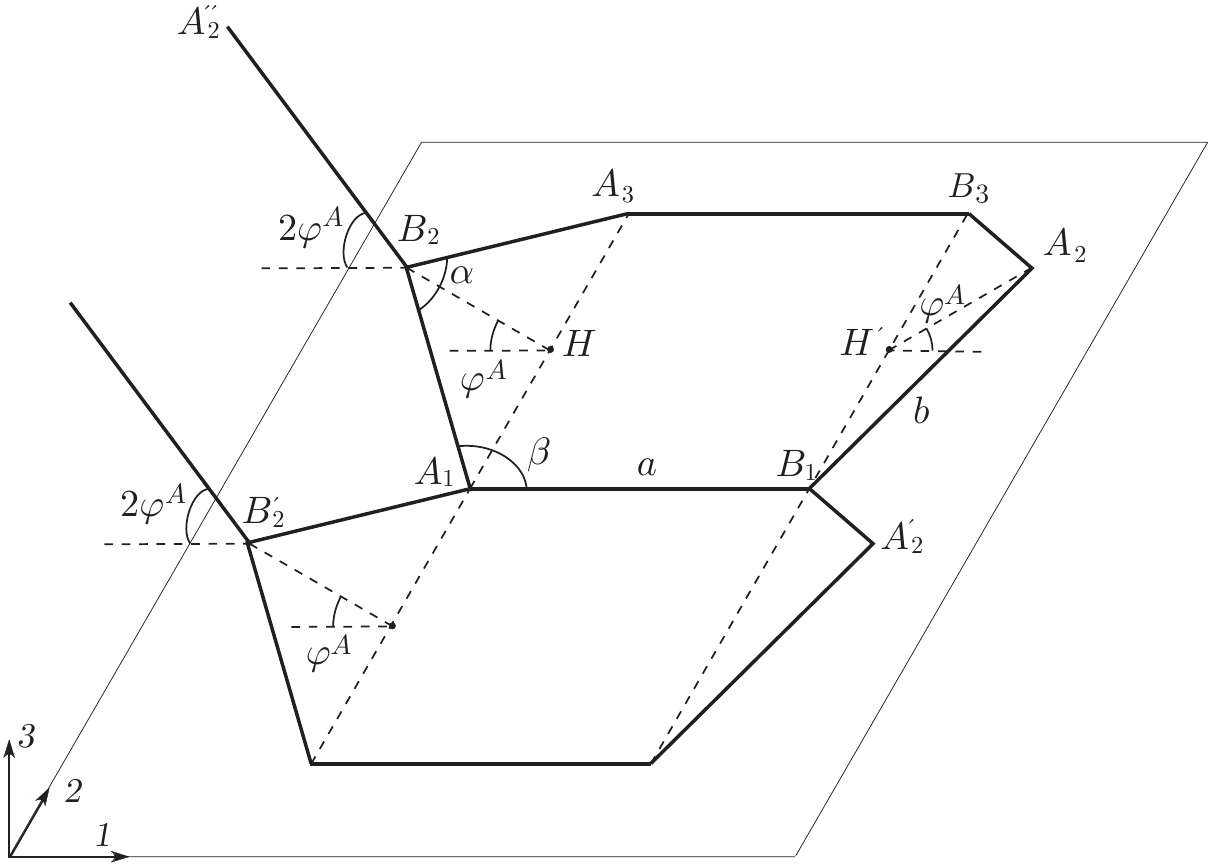}
\caption{The deformed cell of an A-CNT (chiral vector $\hb^A$ parallel to axis 1).}
\label{cell}
\end{figure}

With reference to Fig. \ref{cell},
let $\varphi^{A}$ be the angle between the plane of $A_1 B_1 B_3$ and the plane of $B_1 A_2 B_3$. Since $2\pi=2 n_1\varphi^{A}$, we have that
\begin{equation}\label{eqn1}
\varphi^{A}=\frac{\pi}{n_1}\,.
\end{equation}
%
For geometric compatibility, the bond angles $\alpha$ and $\beta$ must satisfy the following condition:
%
%
\begin{equation}\label{geomcompA}
\cos\beta=-\cos\frac{\alpha}{2}\cos\varphi^A\,,
\end{equation}
whence
\begin{equation}\label{betafunA}
\beta^A=\widetilde{\beta}^{A}(\alpha,\varphi^A):=\arccos\left(-\cos\frac{\alpha}{2}\cos\varphi^{A}\right).
\end{equation}
Moreover, the dihedral angles can be expressed in terms of $\alpha$ and $\beta$ with the use of the following relations:
\begin{equation} \label{gamma1A}
\sin\beta\,\sin\frac{\Theta_1}{2}=\cos\frac{\alpha}{2}\sin\varphi\,,\quad
 \sin\beta\,\sin\Theta_2=\sin\varphi\,, \quad
\Theta_3=2\,\Theta_2\,, \quad \Theta_4= \Theta_5=0\,,
\end{equation}
whence expressions for $\Theta_1^A=\widetilde\Theta_1^A(\alpha,\varphi^A)$ and $\Theta_2^A=\widetilde\Theta_2^A(\alpha,\varphi^A)$ follow, that we here safely omit (see Appendix A1 for details). In conclusion, for an A-CNT of whatever length, the substring \eqref{subs} has the form:
 \begin{equation}\label{subsA}
 \xib_{sub}^A
 =(a,b,\alpha,\widetilde{\beta}^{A}(\alpha,\varphi^A),\widetilde\Theta_1^A(\alpha,\varphi^A),\widetilde\Theta_2^A(\alpha,\varphi^A), 2\widetilde\Theta_2^A(\alpha,\varphi^A),0,0);
 \end{equation}
only three out of nine kinematic variables -- the bond lengths $a,b$ and the bond angle $\alpha$ -- and $n_1$, one of the two size variables, determine the exact configuration.
\subsubsection{Z-CNTs}
\begin{figure}[h]
\centering
\includegraphics[scale=1.5]{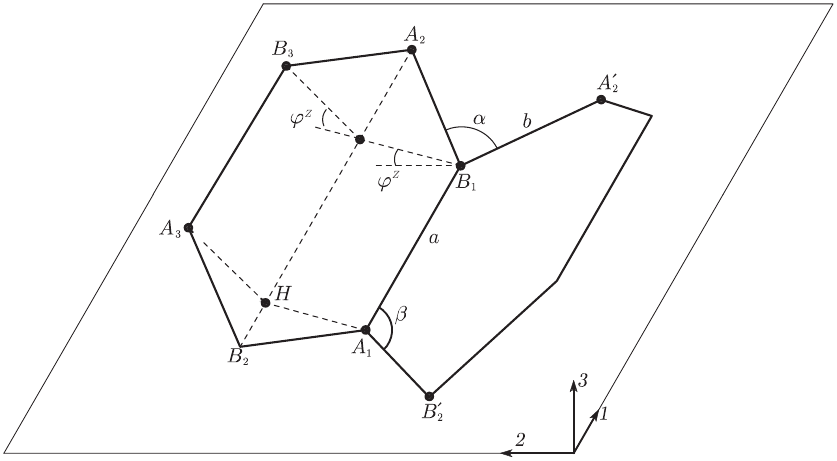}
\caption{The deformed cell of a Z-CNT (chiral vector $\hb^Z$ parallel to axis 2).}
\label{zigzag}
\end{figure}

Consider now  Fig.~\ref{zigzag}, and proceed in parallel to the previous subsection. Then,
the angle $\varphi^{Z}$ between the planes of $A_1 B_1 A_2$ and $A_2 B_3 A_3$ is:
\begin{equation}\label{eqn2}
\varphi^{Z}=\frac{\pi}{n_2}\,;
\end{equation}
the geometric compatibility condition for bond angles is:
\begin{equation}\label{geomcompZ}
\sin\beta\cos\frac{\varphi^{Z}}{2}=\sin\frac{\alpha}{2}\,,
\end{equation}
%
%
whence
\begin{equation}\label{betafunZ}
\beta^Z=\widetilde{\beta}^{Z}(\alpha,\varphi^Z):=\pi-\arcsin\left(\frac{\sin\displaystyle\frac{\alpha}{2}}{\cos\displaystyle\frac{\varphi^{Z}}{2}}\right)\,;
\end{equation}
and, finally,
\begin{equation} \label{gamma1Z}
 \Theta_1=\varphi\,,\quad
 \sin\alpha\sin\Theta_2=\sin\beta\sin\varphi\,, \quad
 \Theta_3=0\,, \quad \Theta_4=2\,\Theta_2\,, \quad \Theta_5=0\,,
\end{equation}
whence the form of function $\Theta_2^Z=\widetilde\Theta_2^Z(\alpha,\varphi^Z)$. In conclusion, for an Z-CNT of whatever length, the substring \eqref{subs} has the form:
 \begin{equation}\label{subsZ}
 \xib_{sub}^Z
 =(a,b,\alpha,\widetilde{\beta}^{Z}(\alpha,\varphi^Z),\varphi^Z,\widetilde\Theta_2^Z(\alpha,\varphi^Z),0,2\,\widetilde\Theta_2^Z(\alpha,\varphi^Z),0);
 \end{equation}
once again, only three kinematic variables and one size variables count to determine the exact configuration, namely, $a,b,\alpha$ and $n_2$.
%


\subsection{Radii and lengths}\label{rl}
%
%
We now show how exact radius and length of an achiral CNT can be computed in terms of the relative $(n_1,n_2)$ pair and the bond-related kinematic variables in the substrings \eqref{subsA} and \eqref{subsZ}.

\subsubsection{A-CNTs}
It is not difficult to see, with the help of Figure~\ref{radii}, that the following geometric compatibility relation holds:
\begin{figure}[h!]
\centering
\includegraphics[scale=0.86]{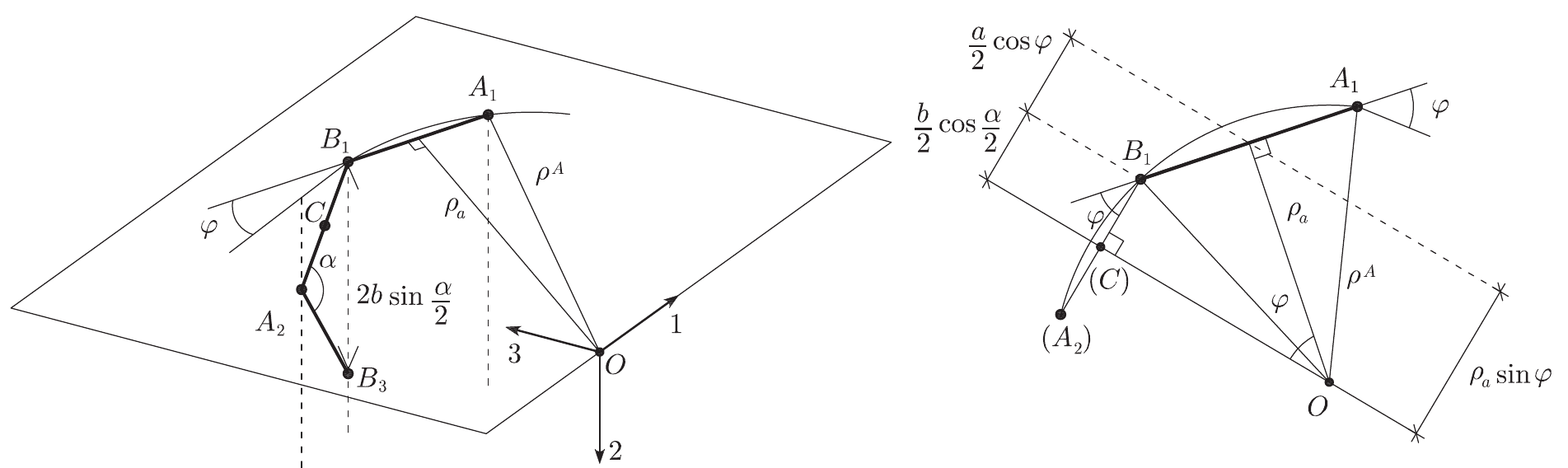}
\caption{Local geometry of an A-CNT.}\label{radii}
\end{figure}
%
%
%
%
\begin{equation}\label{rhoa}
\rho_a\sin\varphi^A=\frac{b}{2}\cos\frac{\alpha}{2}+\frac{a}{2}\cos\varphi^A\,,
\end{equation}
where $\rho_a$ is the distance of the CNT's axis an $a$-type bond. On the other hand,
\begin{equation}\label{RA}
\rho^A=\sqrt{\rho_a^2+\frac{a^2}{4}}\,,
\end{equation}
where $\rho^A$ denotes the cylinder's \emph{exact radius}; consequently, $\rho^A$ too depends only on the kinematic parameters $a,b,\alpha$ and on the size parameter $n_1$, by way of $\varphi^A$.
Figure \ref{radii} is also expedient to see that the exact length of an A-CNT depends as follows from the parameters $b,\alpha$, and $n_2$:
\begin{equation}\label{L2A}
\lambda^A=2\sin\frac{\alpha}{2}\,n_2 \,b\,.
\end{equation}
Formulae \eqref{rhoa}-\eqref{RA} and \eqref{L2A} give the exact dimensions of an A-CNT in terms of the equilibrium bond lengths $a,b$ and bond angle $\alpha$ and of its size parameters $n_1,n_2$.

The exact length given by \eqref{L2A} differs from the nominal length, which is:
\begin{equation}\label{lambdaA0}
\lambda_0^A=\sqrt{3}\,n_2\,r_0;
\end{equation}
we have that
$$
\lambda_0^A/\lambda^A=\frac{\sin\pi/3}{\sin\alpha/2}\,\ r_0/b.
$$
The exact radius must be compared with the nominal radius:
\begin{equation}\label{nomrad}
\rho_0^A=\frac{3}{4\pi}\,n_1r_0
\end{equation}
(cf. the relevant developments in Section \ref{nom}); we expect the former to be slightly larger, an intuitive prediction that our numerical computations generally confirm.
The difference between $\rho^A$ and $\rho_0^A$ becomes negligible for large size indices (e.g., this difference is less than 1\% for a (6,6)-CNT, for which  $\rho^A\simeq0.4$ nm). For large diameters (that is, when $\varphi^A$ is small because $n_1$ is large, due to \eqref{eqn1}), we have that:
\begin{equation}\label{softening}
\rho^A\simeq  \frac{1+b/a\,\cos\alpha/2}{2\pi}n_1\,a;
\end{equation}
accordingly,
\begin{equation}\label{soft}
\rho_0^A/\rho^A\simeq\frac{3}{2(1+b/a\,\cos\alpha/2)}\, r_0/a\,.
\end{equation}
%

\subsubsection{Z-CNTs}
\begin{figure}[h!]
\centering
\includegraphics[scale=1]{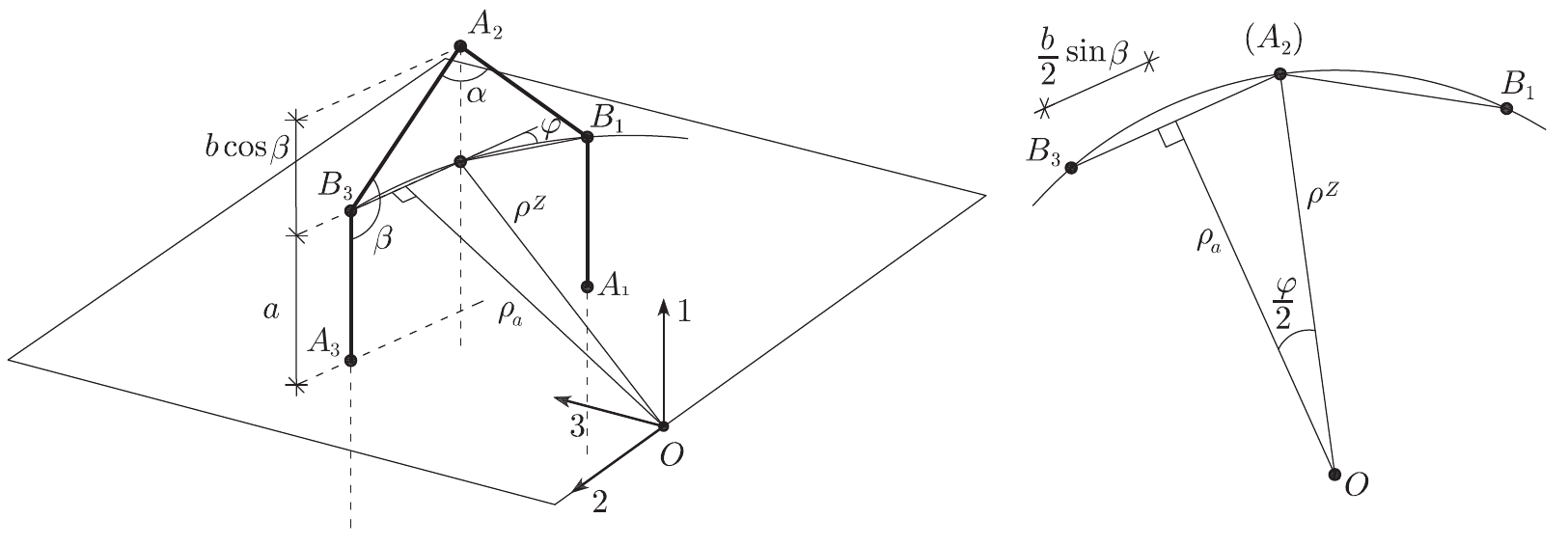}
\caption{Local geometry of a Z-CNT.}\label{rhoZ}
\end{figure}

With the help of Fig. \ref{rhoZ}, it is not difficult to see that the exact radius of a $Z$-CNT is:
\begin{equation}\label{RZ}
\rho^{Z}=\frac{\sin\beta}{2\sin\varphi^Z/2}\,b\,,
\end{equation}
while the nominal radius is
\begin{equation}\label{nomradz}
\rho_0^Z=\frac{\sqrt{3}}{2\pi}\,n_1r_0\,.
\end{equation}
As before, the difference between  $\rho^Z$ and $\rho_0^Z$ is negligible when the size index $n$ is large (e.g., this difference is less than 1\% for a (12,0)-CNT, for which $\rho^Z\simeq 0.5$ nm).

As to the exact length, Fig.  \ref{rhoZ} helps to realize that
\begin{equation}\label{L1Z}
\lambda^{Z}=(1-b/a\cos\beta)n_1\,a\,,
\end{equation}
to be compared with the nominal length:
\begin{equation}\label{lambdaZ0}
\lambda_0^Z=\frac{3}{2}\,n_1\,r_0.
\end{equation}
\section{The self-stress state of CNTs}\label{Equil}
In this section, the bulk of our paper, we  first determine the balance and constitutive equations to be solved for the self-stress state in an achiral CNT, under the assumption that the interactions of C atoms are governed by a REBO interatomic potential (Section \ref{Equil1}). To do the same under the same assumption for chiral CNTs would be exceedingly complicated and, at bottom, of modest importance, given that a numerical minimization of the relative potential (Section \ref{chiral}) yields, in addition to the equilibrium \emph{binding energy}, the equilibrium order-parameter string and hence the corresponding self-stress state, with the use the constitutive equations we derive  in Section \ref{Equil2}.

Whatever the chirality, we ignore \emph{non-bonded interactions}, whose contribution to the self-stress state would be presumably of a lower order of magnitude.\footnote{Non-bonded interactions are usually accounted for by including of a Lennard-Jones term in the interatomic potential \cite{Stuart2000}. Given that the L-J potential is spherical, and that by symmetry all atoms in a CNT are equivalent, the corresponding non-bonded self-stress would be the same as that produced by a small radial force applied on each atom.}
\subsection{Achiral case}\label{Equil1}
The main outcome of Section \ref{OPS} is that both order-parameter substrings \eqref{subsA} and \eqref{subsZ} depend solely on the three independent Lagrangian coordinates $a,b$, and $\alpha$, and that, moreover, their last entry $\Theta_5$ is null. Our first goal in this section is to specify what mapping $$(a,b,\alpha)=\qb\mapsto V=\widehat V(\qb)$$ of the type introduced in Section \ref{natequi} is to be made stationary to find natural equilibria.

We begin by counting, type by type, the number of bond lengths, bond angles, and dihedral angles, that take equal values in a natural equilibrium configuration of an achiral CNT with a  $(n_1,n_2)$ parent strip as in Fig. \ref{rectangle}. We find, respectively,
\begin{equation}\label{ns}
n_a = n_1 n_2,\quad n_b = n_\alpha = 2 n_1 n_2,\quad n_\beta = 4 n_1 n_2,\quad
n_{\Theta_1} = 2 n_a,\quad n_{\Theta_2} = 2 n_b,\quad n_{\Theta_3} = n_{\Theta_4} =  n_b\,.
\end{equation}
Now, as anticipated at the end of Section \ref{geokin}, the total potential of an achiral CNT in equilibrium can be written in terms of a no-redundancy sum over diatomic bonds;  according to the 2nd-generation REBO potential we use (see Appendix B), a specific set of bond and dihedral angles is associated to each of the two types of diatomic bonds we distinguished. In view of this state of affairs, we write the total potential $V$ as follows:
\begin{equation}\label{V:AZ}
V=n_a V_a + n_b V_b = n_1 n_2 (V_a+2V_b)\,,
\end{equation}
where
\begin{equation}\label{Va:AZ}
\begin{aligned}
V_a(a,\beta,\Theta_1)&=V_R(a)+b_a(\beta, \Theta_1)\,V_A(a)\,,\\
V_b(b,\alpha,\beta,\Theta_2,\Theta_3,\Theta_4)&=V_R(b) +
b_b(\alpha, \beta, \Theta_2, \Theta_3, \Theta_4)\,V_A(b)\,.
\end{aligned}
\end{equation}
The forms of the \emph{attraction} and \emph{repulsion} functions $V_A$ and $V_R$ and of the \emph{bond-order} functions $b_a$ and $b_b$ are found in  Appendix B.2, equations \eqref{bab}; here, it is sufficient to know that they all are as smooth as needed to justify our further developments, and to warn the reader that the entries $(\beta, \Theta_1)$ of $b_a$  and $(\beta,\Theta_2, \Theta_3, \Theta_4)$ of $b_b$ must be thought of as depending  either on $\alpha$ and $\varphi^A$ as specified by \eqref{betafunA} and \eqref{gamma1A} or on  $\alpha$ and $\varphi^Z$ as specified by \eqref{betafunZ} and \eqref{gamma1Z}. With slight abuse of the notation introduced in Section \ref{natequi}, we set
\begin{equation}\label{achipot}
V=\widetilde V(\xib):=n_aV_a(a,\beta,\Theta_1)+n_bV_b(b,\alpha,\beta,\Theta_2,\Theta_3,\Theta_4),\quad\xib:=(a,b,\alpha,\beta,\Theta_1,\Theta_2,\Theta_3,\Theta_4),
\end{equation}

We are now in a position to write the stationarity condition of the potential $\widehat V$. We do it in a form involving the stress mapping $\widetilde\chib$:
\[
\delta V=\chib\cdot\delta\xib=0,\quad \chib=\widetilde \chib(\xib):=\partial_{\xib}\widetilde V\,,
\]
(cf. \eqref{increV}).
Proceeding as in Section \ref{natequi}, this stationarity condition can be re-written equivalently in a form involving also the equilibrium operator $\Ab=(\partial_\qb\xib)^T$:
\begin{equation}\label{eccoqui}
\Ab\chib=\0
\end{equation}
(cf. \eqref{ecco}). Now, the matrix form of $\Ab$ is:
\begin{equation}\label{A}
\left[\begin{array}{c}\Ab\end{array}\right]=\left[\begin{array}{cccccccc}
1 & 0 & 0 & 0 & 0 & 0 & 0 & 0 \\
0 & 1 & 0 & 0 & 0 & 0 & 0 & 0 \\
0 & 0 & 1 & \beta,_\alpha & \Theta_1,_\alpha & \Theta_2,_\alpha & \Theta_3,_\alpha & \Theta_4,_\alpha
\end{array}\right]\,;
\end{equation}
a stress-mapping string $\chib$ consists of the following entries:
\begin{equation}\label{chi}
\left[\begin{array}{c}\chib\end{array}\right]=\left[\begin{array}{c}
n_a\,\sigma_a \\
n_b\,\sigma_b  \\
n_\alpha\,\tau_\alpha\\
 n_\beta\,\tau_\beta\\
 n_{\Theta_1} \, \Tc_1\\
 n_{\Theta_2}  \, \Tc_2\\
 n_{\Theta_3}   \, \Tc_3\\
 n_{\Theta_4}   \, \Tc_4\\
\end{array}\right]\,.
\end{equation}
%
%
%
With the use of \eqref{ns}, the equilibrium equation \eqref{eccoqui} becomes:
\begin{equation}\label{finali}
\left\{\begin{array}{c}\sigma_a= 0\,,\quad
\sigma_b= 0\,, \\
\tau_\alpha+2\beta,_\alpha\, \tau_\beta +\sum_{i=1}^4 \,\Theta_i,_\alpha\,\Tc_i= 0\,.
\end{array}\right.
\end{equation}
As a glance to \eqref{A} reveals, the nullspace of $\Ab$ is five-dimensional; this means that an achiral CNT can sustain five independent self-stress states, one work-conjugated to bond-angle changes, the other four with dihedral-angle changes, and all of them having $\sigma_a=\sigma_b=0$.

We term all of $\sigma_a, \sigma_b, \tau_\alpha, \tau_\beta$, and $\Tc_i$,  {\em nanostresses},  work-conjugate to changes of, respectively, bond lengths, bond angles, and dihedral angles. Here is how individual nanostresses depend on order-parameter strings:
\begin{equation}\label{nanostress}
\begin{aligned}
\sigma_a & = V_R'(a)+b_a(\beta,\Theta_1)\,V_A'(a)\,,\quad
\sigma_b  = V_R'(b)+b_b(\alpha,\beta,\Theta_2,\Theta_3,\Theta_4)\,V_A'(b)\,,\\
\tau_\alpha & = b_b,_\alpha\!(\alpha,\beta,\Theta_2,\Theta_3,\Theta_4)\, V_A(b) \,,\quad
\tau_\beta  = \frac{1}{4}\big(b_a,_\beta\!(\beta,\Theta_1) \,V_A(a)+ 2 b_b,_\beta\!(\alpha,\beta,\Theta_2,\Theta_3,\Theta_4) \,V_A(b)\big)\,,
\\
\Tc_1 & = \frac{1}{2}\,b_a,_{\Theta_1}\!(\beta,\Theta_1) \,V_A(a)\,,\quad
\Tc_2  =  \frac{1}{2}\,b_b,_{\Theta_2}\!(\alpha,\beta,\Theta_2,\Theta_3,\Theta_4) \,V_A(b)\,,\\
\Tc_3 & = b_b,_{\Theta_3}\!(\alpha,\beta,\Theta_2,\Theta_3,\Theta_4)\, V_A(b)\,,\quad
\Tc_4 = b_b,_{\Theta_4}\!(\alpha,\beta,\Theta_2,\Theta_3,\Theta_4) \,V_A(b)\,.
\end{aligned}
\end{equation}
Remarkably, both the balance equations \eqref{finali} and the  constitutive equations \eqref{nanostress} are valid whatever REBO potential one chooses, no matter whether it is Tersoff's or Brenner's (of first or second generation), or others.

It is important to realize that the solution of the nonlinear system \eqref{finali}-\eqref{nanostress} depend in general not only on the type of the achiral CNT under attention but also on its size, because functions $\widetilde\beta$ and $\widetilde\Theta_i$ do (cf. \eqref{subsA} and \eqref{subsZ}). Here is how the third of \eqref{finali} depends on which of the two achiral CNTs is dealt with:
\begin{equation}\label{mairichiamata}
\begin{aligned}
\tau_\alpha^{A}+2\,\beta^{A},_\alpha\, \tau_\beta^{A} +\sum_{i=1}^3 \,\Theta_i^{A},_\alpha\,\Tc_i^{A}&= 0\,,\\
\tau_\alpha^{Z}+2\,\beta^{Z},_\alpha\, \tau_\beta^{Z} +{\Theta_2^{Z}},_\alpha\Tc_2^{Z}+{\Theta_4^{Z}},_\alpha\Tc_4^{Z}&= 0\,;
\end{aligned}
\end{equation}
the expressions of the derivatives $\beta^N,_\alpha$ and ${\Theta^N_i},_\alpha$ (where $N=A$ or $Z$ and $i=1,\ldots,4$), are found in Appendix A, equations \eqref{betader} and \eqref{betaderZ}.  We believe that, for each type and whatever the size, there is only one natural solution $\qb_0^N=(a_0^N,b_0^N,\alpha_0^N)$; our belief is substantiated by the outcome of the numerical procedure we use to determine $\qb^N$:
\vskip 3pt
\noindent {\sc Step 1 -} compute the nominal set $\qb^N_0(n)$ of an achiral CNT of very large size index $n$, whose nominal and natural diameters almost coincide no matter how computed (by means of DFT or TB techniques, say);

\noindent {\sc Step 2 -} solve numerically  system \eqref{finali} for $\qb^N(n)$,  \emph{with $\qb^N_0(n)$ as initial guess};

\noindent {\sc Step 3 -} with  $\qb^N(n)$ as initial guess, solve system \eqref{finali} for a CNT of the same type  and smaller radius.
\vskip 3pt
\noindent Step 3 is to be iterated  as many times as desired and possible (we have decreased the size index until $n=3$ in the A case, until $n=5$ in  the Z case).\footnote{The numerical routine we employ is the one available in the software MATLAB (routine {\em fsolve}). This routine is a realization of the so-called \emph{Trust-Region Method}, according to which at each step an approximation of the objective function (namely, the sum of squares of the residues) is minimized over a region whose size is adjusted to improve convergence speed.
}

Needless to say, given $\qb^N$, the natural radius and length are computable with the formulae derived in Section \ref{rl}, the self-stress state with the use of \eqref{chi} and \eqref{nanostress}.

{\remark {\em
The three types of nanostresses we consider, $\sigma$, $\tau$ and $\mathcal{T}$, are -- we repeat -- work-conjugate to changes in, respectively,  bond lengths, bond angles, and dihedral angles. Were CNTs be visualized as discrete mechanical structures consisting of pin-jointed sticks, those nanostresses would be associated with the response to structural deformations of a set of axial, rotational, and dihedral springs. An order-parameter substring, together with the collection of nanostresses associated to it by the stress mapping, yields the information necessary to evaluate the energy density  per lattice cell, opening the way to the use of homogeneization techniques \cite{Chen_1998,Davini_2011,Davini_2014}. In this connection, it is worth mentioning that in \cite{Chen_1998,Liu_2009} a couple-stress continuum is regarded to be the homogenized version of discrete mechanical structures of the above type, in the absence of dihedral springs.  It remains to be seen, were dihedral springs included, what higher-gradient elasticity model would turn out to be the appropriate continuum limit.
}}

\subsection{Chiral case}\label{Equil2}

\begin{figure}[h]
\centering
\includegraphics[scale=1]{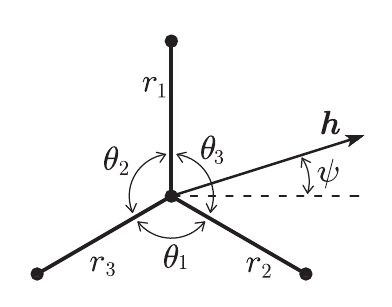}
\label{assi}
\caption{The bond unit of a chiral CNT.}
\end{figure}

 To see why the chiral case is much more complicated to deal with than the achiral case, it is expedient to begin by contrasting the respective order-parameter substrings. In the former case, we have:
\begin{equation}\label{subchi}
\csib_{sub}^{c}=(r_1,r_2,r_3,\theta_1,\theta_2,\theta_3,\Theta_{11},
\Theta_{12},\Theta_{13},\Theta_{21},\Theta_{22},\Theta_{23},
\Theta_{31},\Theta_{32},\Theta_{33})\,,
\end{equation}
where $r_i$, $\theta_i$ are the typical bond lengths and bond angles, respectively, and $\Theta_{ij}$ is the $j$-th dihedral angle associated to the $i$-th bond (see Fig.\,\ref{assi};
for each type of bond, there are three types of dihedral angles, because, by symmetry, two of the four dihedral angles associated to a bond are equivalent to each other). In the achiral case, there are only two distinct bond lengths $a,b$ and bond angles $\alpha,\beta$, and only four nonnull dihedral angles $\Theta_1,\ldots,\Theta_4$:
 \begin{equation}\label{subss}
 \xib_{sub}^{ac}=(a,b,\alpha,\beta,\Theta_1,\ldots,\Theta_4)
 \end{equation}
(cf. \eqref{subs} and the second of \eqref{achipot}; recall that the fifth dihedral angle is always null).

%
For the interatomic potential of a chiral bond unit we take:
%
\begin{equation}\label{Vchi}
V=\widetilde V^{c}(\csib^{c}):=\sum_{i=1}^3\, \widetilde V_i^{c}(\csib^{c}),\quad \widetilde V_i^{c}=V_R(r_i)+ b_i(\theta_{i+1},\theta_{i+2},\Theta_{i1},\Theta_{i2},\Theta_{i3})\,V_A(r_i)
\end{equation}
where functions $V_A$, $V_R$, and $b_i$, are defined in Appendix B.2 (here and henceforth subscripts are to be chosen \emph{modulo} 3). The associated constitutive equations are:
\begin{equation}\label{nnstresschi}
\begin{aligned}
&\sigma_i= V'_R(r_i)+ b_i\,V'_A(r_i)\,,\\
&\tau_i = \frac{1}{2}\,\big((b_{i+1}),_{\theta i}\,V_A(r_{i+1}) + (b_{i+2}),_{\theta i}\,V_A(r_{i+2})\big)\,,\\
&\Tc_{i1} = \frac{1}{2}\,b_{i},_{\Theta {i1}}\,V_A(r_{i})\,,\quad \Tc_{ij} = b_{i},_{\Theta {ij}}\,V_A(r_{i})\,,\quad j=2,3\,.
\end{aligned}
\end{equation}

The mere fact that there are fifteen entries in \eqref{subchi} makes the  writing of stationarity conditions cumbersome and, at bottom, inessential: a direct minimisation of the total interatomic potential yields the equilibrium values of the kinematic variables, with which, on making use of the appropriate constitutive equations, the equilibrium nanostresses, and hence the equilibrium self-stress state, can be evaluated; this we do in Section \ref{chiral}. However, a stationarity condition for the  interatomic potential is particularly easy to find, the one obtained by considering variations describing homothetic expansions of bond units, that is to say, variations such that all bond lengths $r_i$ are equally rescaled, while all bond and dihedral angles are left unchanged. One finds:
%
%
\begin{equation}\label{chieql}
\sum_{i=1}^3\sigma_i\, r_i=0\,,
\end{equation}
with $\sigma_i\;\,(i=1,2,3)$ the bond-length nanostresses. Equation \eqref{chieql} implies that the $\sigma_i$'s cannot all have the same sign; they may be different from zero, though, at variance with the achiral case, in which independent variations of the bond lengths $a,b$ are geometrically compatible.



\section{Results}\label{discuss}
In this section we collect our results and compare them with those in the literature. In all of Fig.s \ref{d}-\ref{nanostresschi} the abscissa is the \emph{natural curvature}, that is, the inverse of natural radius; the data we use to draw the plots are collected in Appendix D, alongside with a selection of similar data taken from the literature.  The first three subsections are dedicated to achiral CNTs and FGSs, the fourth and last to chiral CNTs.

\subsection{Natural \emph{vs.} nominal geometry}
\begin{figure}[h]
\centering
\includegraphics[scale=0.77]{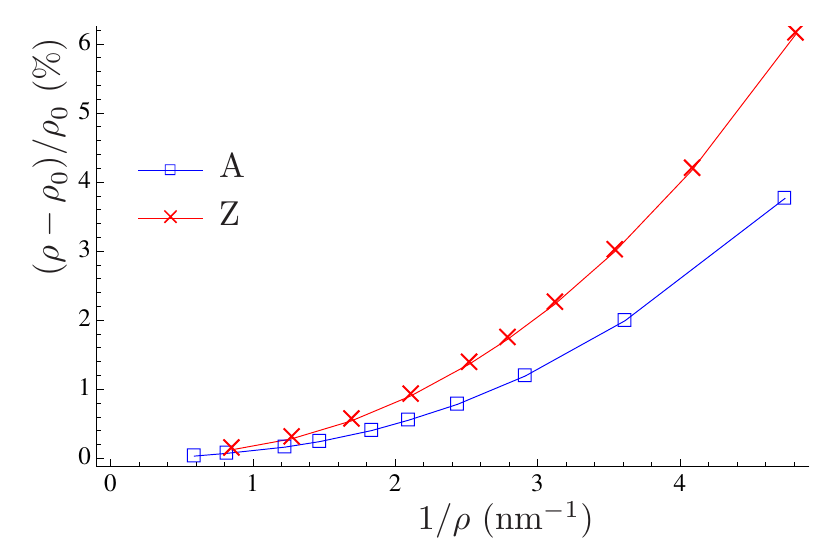}
\qquad
\includegraphics[scale=0.77]{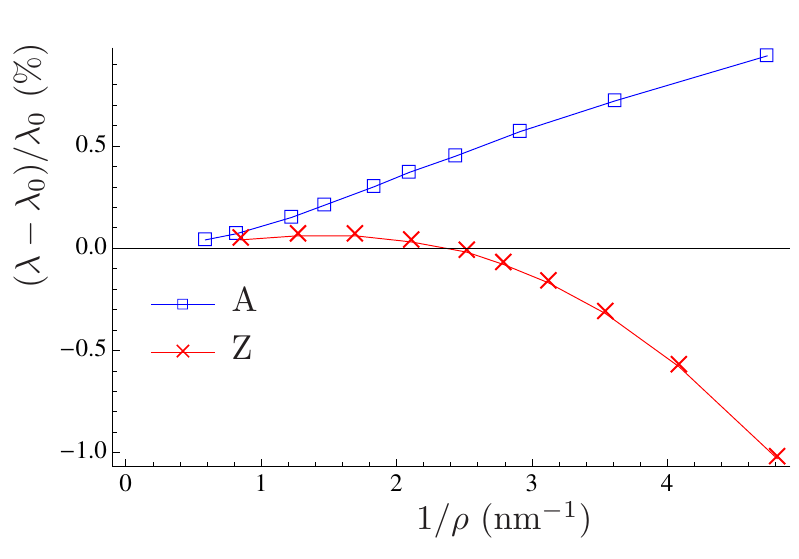}
\caption{Specific difference between natural and nominal radii \emph{vs.} natural curvature (left). Specific difference between natural and nominal lengths \emph{vs.} natural curvature (right).}
\label{d}
\end{figure}

The percentage of the specific difference between natural radius  $\rho$ and nominal radius  $\rho_0$ is plotted in Fig. \ref{d} (left) as a function of the natural curvature $1/\rho$.
We see that the smaller the radius, the larger the difference, a result of ours that confirms all previous calculations \cite{Kanamitsu2002,Machon2002,Kurti2003,Cabria2003,Budyka_2005,Demichelis_2011,Popov2004,Jiang_2003,Jindal_2008}; the natural radii we and others computed for a few small achiral CNTs are listed in Table  \ref{tab:r:comparison}.
\begin{table}[H]
\begin{center}
\caption{Natural radii (nm) of small CNTs.}
\label{tab:r:comparison}
\begin{tabular}{cccccccc}
\ & \ &  \ & \ &  \ & \ & \ & Jindal \\
$(n,m)$ & present &  Mach\'on & Cabria &  Popov & Budyka & Jiang & \& Imtani\\
\ & work &  et al. 2002 & et al. 2003 &   2004 & et al. 2005 & et al. 2003 & 2008 \\
\ & (2nd-g. Brenner) & (DFT) & (DFT) & (TB) & (DFT) & (1st-g. Brenner) & (Tersoff) \\[3pt]
\hline
(3,3) & 0.211 &  0.210 & 0.212 & 0.212 & -     & -     & 0.211 \\
(4,4) & 0.277 &  -     & -     & -     & 0.277 & 0.280 & -     \\
(5,0) & 0.208 &  0.204 & 0.206 & 0.205 & -     & 0.208 & -     \\
\hline
\end{tabular}
\end{center}
\end{table}
\noindent We also see from Figure \ref{d} (right) that the specific difference between natural and nominal lengths is always positive for A-CNTs, while for Z-CNTs it is negative for small radii, again positive and close to zero for large radii.

\begin{figure}[h]
\centering
\includegraphics[scale=0.77]{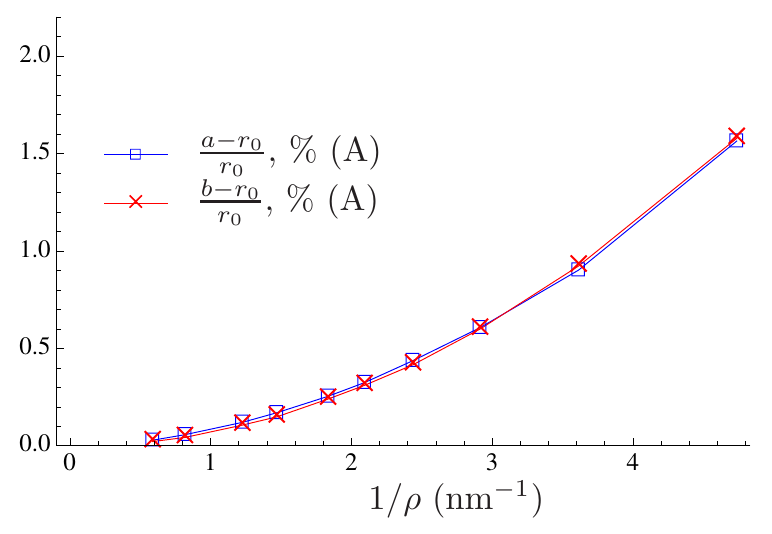}
\quad
 \includegraphics[scale=0.77]{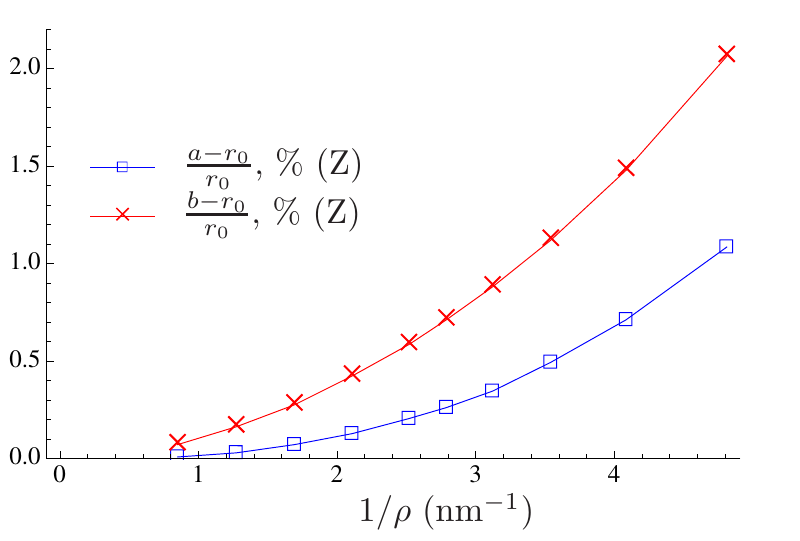}
\caption{
Bond-length specific differences \emph{vs.} curvature, for A-CNTs (left) and Z-CNTs (right).
}
\label{abA}
\end{figure}

In a FGS, the length of a C-C bond, when computed \emph{via} the interatomic potential we use in this study, is $r_0=0.14204$ nm (needless to say, bond angles measure $120^\circ$, and dihedral angles are null).
For single-walled CNTs, \emph{natural bond lengths} differ from their nominal length $r_0$, more and more when radii decrease. Fig. \ref{abA}
%
shows the percentage of the specific difference between natural and nominal lengths of type $a$ and type $b$ bonds as a function of natural curvature, for A- and Z-CNTs. The results for A-CNTs are consistent with those in \cite{Kurti2003}; however, for Z-CNTs, DFT calculations shows that $a$-type bonds are shorter than $r_0$ \cite{Kanamitsu2002,Kurti2003}.

\begin{figure}[H]
\centering
\includegraphics[scale=0.77]{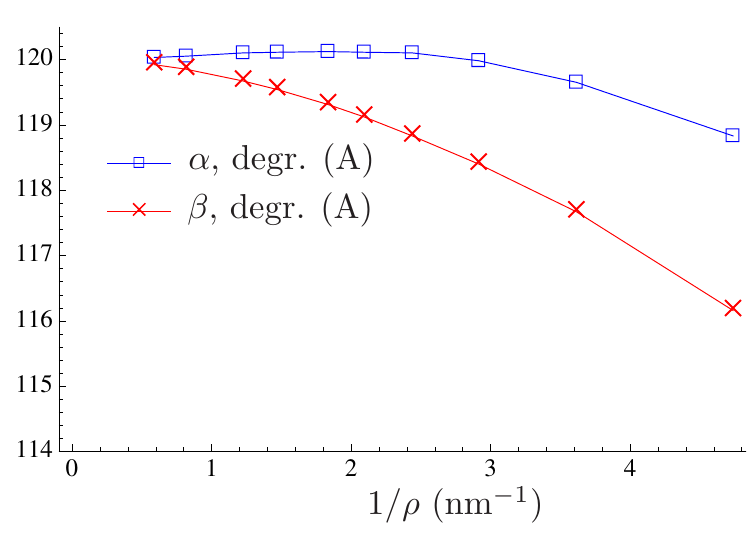}
\qquad
\includegraphics[scale=0.77]{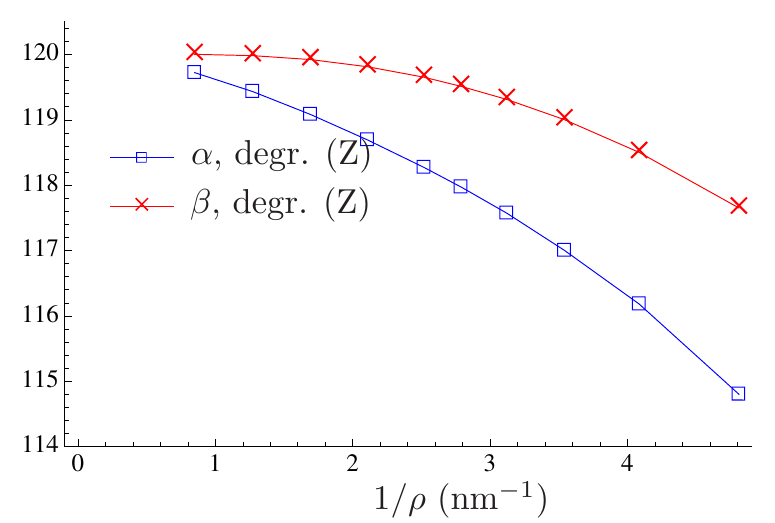}
\\[10pt]
\includegraphics[scale=0.77]{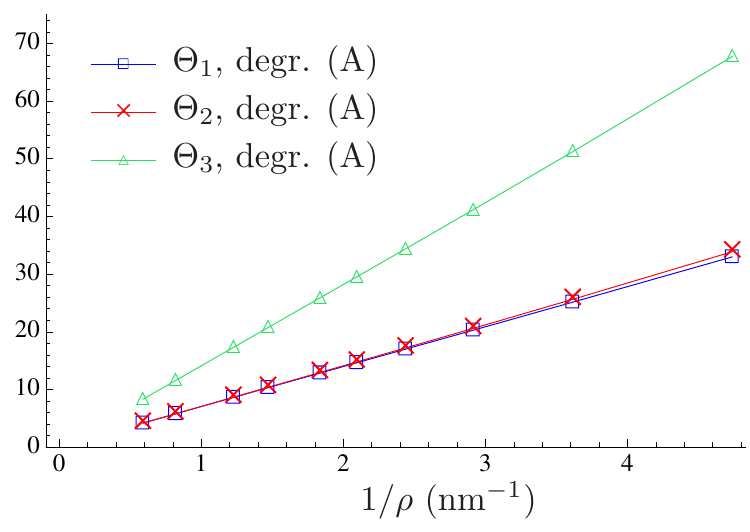}
\qquad
\includegraphics[scale=0.77]{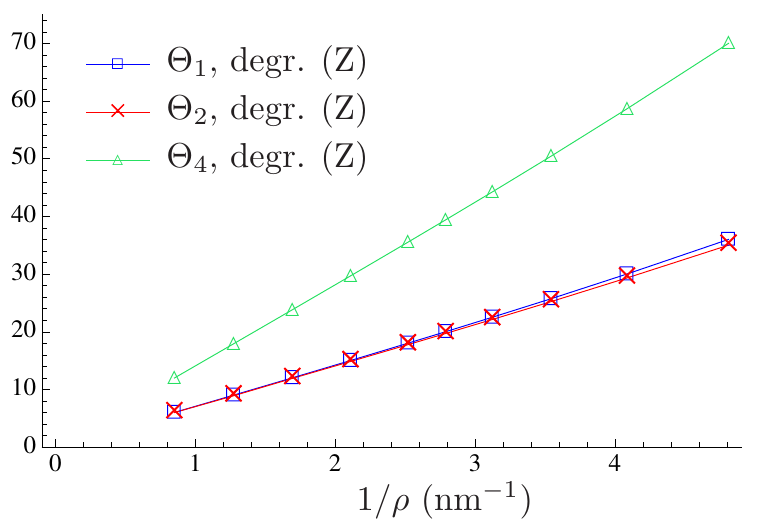}
\caption{
(top) Bond angles $\alpha$ and $\beta$  \emph{vs.} curvature, in A-CNTs (left) and Z-CNTs (right).
(bottom) Dihedral angles $\Theta_1$, $\Theta_2$, $\Theta_3$ \emph{vs.} curvature, in A-CNTs ( left);
 dihedral angles $\Theta_1$, $\Theta_2$, $\Theta_4$ \emph{vs.} curvature, in Z-CNTs (right).
}
\label{abAbis}
\end{figure}

The \emph{natural bond angles} $\alpha$ and $\beta$ we calculated are plotted in Fig.\,\ref{abAbis} (top); we see that they are smaller than $120^\circ$, except for $\alpha$ in the A case, which takes slightly higher values at curvatures smaller than $3$ nm$^{-1}$. We also see that $\alpha$ decreases more in the Z case, and that $\beta$ decreases more in the A case, as expected on the basis of the RUM model. Fig.\,\ref{abAbis} (bottom) shows how the \emph{natural dihedral angles} $\Theta_i$, $i=1,\ldots,4$ behave when the natural curvature varies; note that, as required by, respectively, $\eqref{gamma1A}_3$ and $\eqref{gamma1Z}_4$, $\Theta_3=2\Theta_2$ in the A-case and $\Theta_4=2\Theta_2$ in the Z-case. 

Detailed information about all  hexagonal-lattice parameters is collected in Appendix D, Tables \ref{tab:geometry} and \ref{tab:dh:angles}.
By comparison with the results reported in \cite{Jiang_2003}, where a 1st-generation REBO potential was used \cite{Brenner_1990} (see Table \ref{Jiang:res} in Appendix D), we see that the CNTs' radii are always larger than those we computed, partly due to the fact that the potential employed  in \cite{Jiang_2003} yields ${\widetilde r}_0=0.14507$ nm for the natural bond length in graphene; anyway, in \cite{Jiang_2003}  bond lengths are always larger than ${\widetilde r}_0$, a result qualitatively identical to ours. Notice that in \cite{Jindal_2008}, where Tersoff potential \cite{Tersoff_1988} is employed for studying A-CNTs, $b$-type bonds are shorter than ${\widetilde r}_0$, a result which disagrees with the DFT calculations in \cite{Kurti2003}.
%

\subsection{The self-stress state of achiral CNTs and FGSs}\label{SSS}
\begin{figure}[h]
\centering
\includegraphics[scale=0.77]{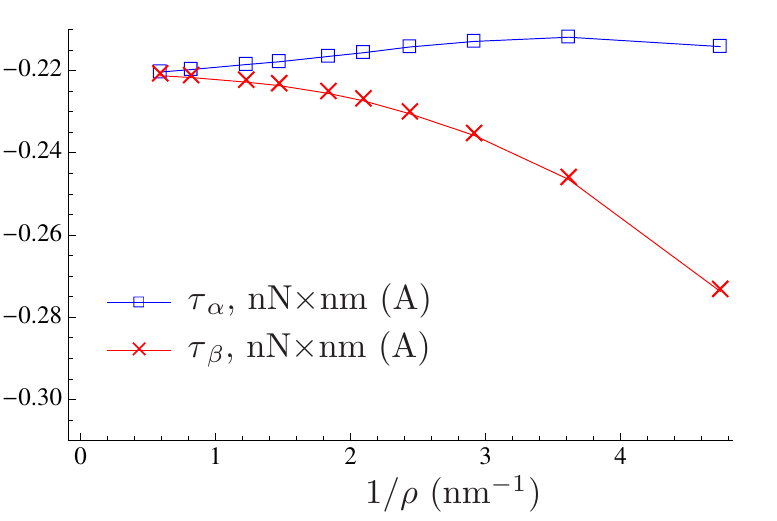}
\qquad
\includegraphics[scale=0.77]{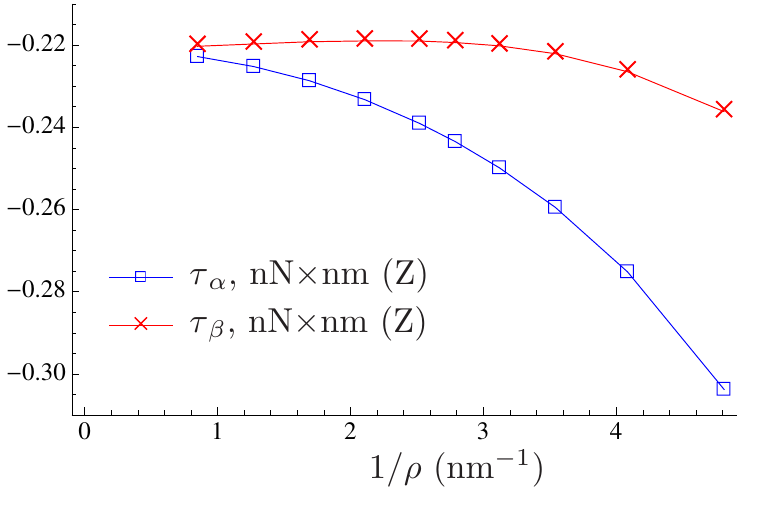}
\\[10pt]
\includegraphics[scale=0.77]{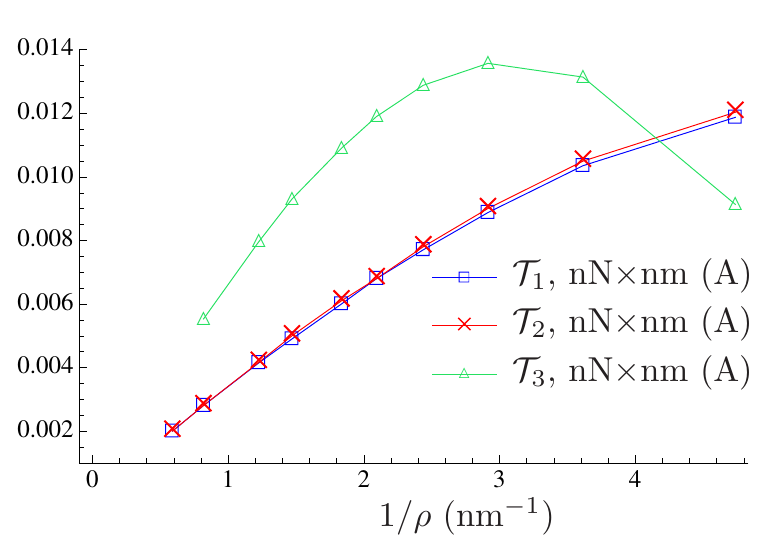}
\qquad
\includegraphics[scale=0.77]{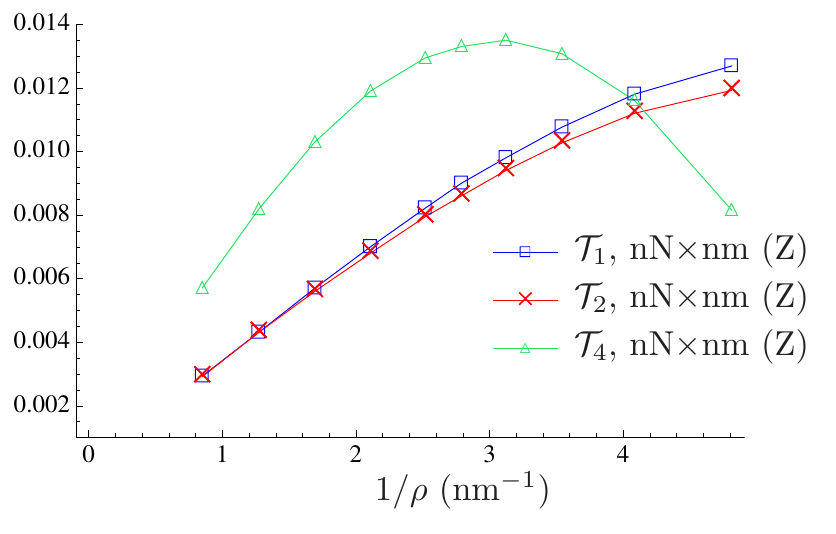}
\caption{(top) Bond-angle self-stresses $\tau_\alpha$ and $\tau_\beta$ \emph{vs.} curvature in A-CNTs (left) and Z-CNTs (right). (bottom) Dihedral-angle self-stresses $\Tc_1$, $\Tc_2$, $\Tc_3$  \emph{vs.} curvature in A-CNTs (left); dihedral-angle self-stresses $\Tc_1$, $\Tc_2$, $\Tc_4$  \emph{vs.} curvature  in Z-CNTs (right).}\label{nanostressA}
\end{figure}

%
Fig.\,\ref{nanostressA} shows how \emph{bond-angle nanostresses} $\tau_\alpha$ and $\tau_\beta$ (top)  and \emph{dihedral nanostresses} $\Tc_i$ (bottom) depend on natural curvature of CNTs (see Table \ref{tab:nanostress}, Appendix D, for the relative numerical information).
%
On setting $a=b=r_0$, $\alpha=\beta=2\pi/3$, and $\Theta_i=0$ ($i=1,\ldots,4$), in relations \eqref{nanostress}, we see that, while the dihedral nanostress is null, the bond-angle nanostress takes the value $\tau_g = -0.2209$ nN$\times$nm: hence,   \emph{FGSs too are stressed in their natural state}. 

To provide evidence  that the self-stress state of FGSs is far from being negligible, we made use of the developments in Appendix C and showed that a traction load in the direction of $\hb^Z$ ($\hb^A$) of $30$ nN/nm   -- some 3/4 of the failure load, according to \cite{Lee2008} --  induces variations of $\tau_\alpha$ and $\tau_\beta$ that are equal, respectively, to $-77\%$ and $+18\%$ ($+53\%$ and $-49\%$) of  $\tau_g$, their identical value prior to load application. It is worth noticing that a use of 1st-generation potentials \cite{Tersoff_1988,Tersoff_1989,Brenner_1990} would lead to overestimating $\tau_g$, in spite of the fact that those potentials were calibrated with reference to a collection of physical and chemical properties quite similar to that used for the 2nd-generation Brenner potential we adopted: according to our computations, the value of $\tau_g$ is $-0.3253$ nN$\times$nm for the Tersoff potential \cite{Tersoff_1988,Tersoff_1989}, and $-0.2554$ nN$\times$nm for the 1st-generation Brenner potential \cite{Brenner_1990}.  Needless to say, a re-parametrization of these 1st-generation potential intended to achieve a fair prediction of the self-stress state is likely to offset the predictions of other kinds of mechanical properties. 

{\remark{\em
In \cite{Xu2013}, graphene's \emph{bending stiffness} is presumed to be related to a bond-angle nanostress, and it is argued that, if such a nanostress were actually present, then   
 the bending stiffness would be equal to $\tau_g/2$, that is, the value the bending stiffness should have in the absence of a dihedral contribution, according to the computations reported in \cite{Lu_2009}.}
 }

\subsection{Roll-up energy}\label{rollupenergy}
We call
\emph{roll-up energy}  the difference in energy per atom of a CNT  and its `parent' FGS, in their respective natural configurations; the energy of the latter is equal to $-7.3951$ eV/atom, when estimated by a 2nd-generation Brenner potential.

\begin{figure}[H]
\centering
\includegraphics[scale=0.77]{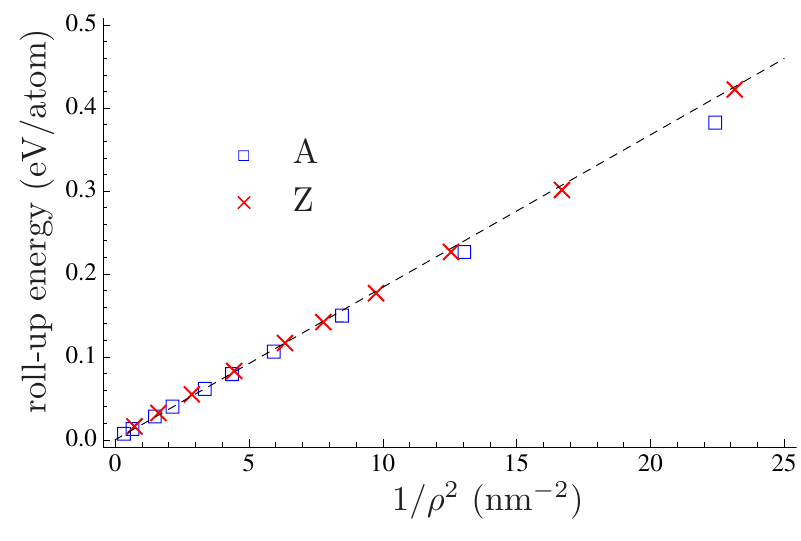}\qquad
\includegraphics[scale=0.77]{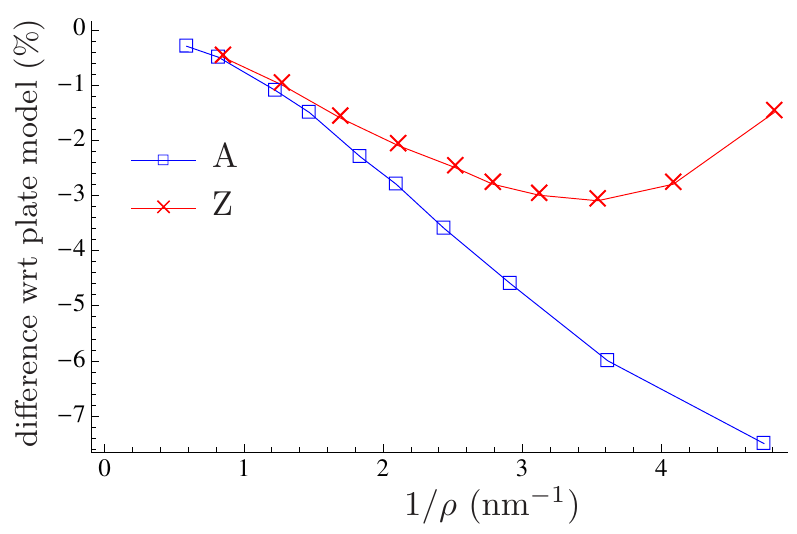}
    \caption{(left) Roll-up energy \emph{vs.} natural curvature squared (the dashed line is drawn on taking  $D=0.03675$\,eV\,nm$^2$/atom in the `thin-plate formula' $\frac{1}{2}{D}\rho^{-2}$). (right)  Percentage difference of roll-up and `thin-plate' plate energies \emph{vs.}  natural curvature.}\label{energyvsd2}
\end{figure}

Our findings about roll-up energy are plotted in Fig. \ref{energyvsd2} (left), as functions of the natural curvature squared,
for
A-CNTs whose size index ranges from 3 to 25 and for Z-CNTs whose size index ranging from 5 to 30 (the corresponding radii fall in the $(0.208,1.696)$ nm interval). Our choice of abscissa is motivated by the fact that the roll-up energy is often evaluated by a formula, $\frac{1}{2}{D}\rho^{-2}$, which imitates the formula for a bent linearly elastic thin plate, and consistently called \emph{folding energy}; the value of the stiffness constant $D$, which depends on the material and the third power of the thickness  in the case a thin plate,  is taken equal to $0.03675$\,eV\,nm$^2$/atom \cite{Lu_2009}. Of course,  as some DFT and TB calculations confirm  \cite{Robertson1992,LuJ1997,Hernandez1998,Sanchez1999,Kanamitsu2002,Kurti2003}, the above `thin-plate formula' is less and less reliable as a CNT's curvature grows big; for us, at variance with what is reported in  \cite{Kurti2003}, that formula always provides an estimate from above of the roll-up energy (see Fig. \ref{energyvsd2} (right)).\footnote{ Although CNTs are never manufactured by rolling FGSs up, this ideal procedure is sometimes evoked to to justify the dependence of the folding energy on $\rho^{-2}$ and, more generally, to motivate the
 building of continuum theories for CNTs and FGs.  In \cite{Luhuang2009}, a non-linear continuum model for graphene is proposed; the bending of an FGS into a cylindrical CNT of given radius is regarded as the outcome of a geometrically defined process and not of an energy minimization; the final rolled-up configuration is assumed to be stressed (although,  in principle, energy minimization might have produced an unstressed configuration) and serves as a reference for further mechanical deformations, this time regarded as minimizers of a total energy functional to be found under the constraint that the CNT's radius stays fixed. In the same spirit, a finite-element model for CNTs was developed in \cite{Pantano2004}, where the self-stress state is computed by means of the classical linear shell theory, on the assumption that the energy stored because of the roll-up procedure is $\frac{Et^3}{24\rho^2(1-\nu^2)}$, with $E$ the Young modulus, $\nu$ the Poisson ratio and $t$ the wall thickness. If not for other reasons, adopting model equations from the linear theory of thin elastic structures where an evaluation of an elusive quantity like structure thickness is essential seems to us  questionable (see \cite{Bajaj_2013} and \cite{Huang2006} for a discussion about the notion of thickness in continuum models for CNTs).
 }

\begin{figure}[h]
\centering
\begin{minipage}[c]{.40\textwidth}
\includegraphics[scale=0.7]{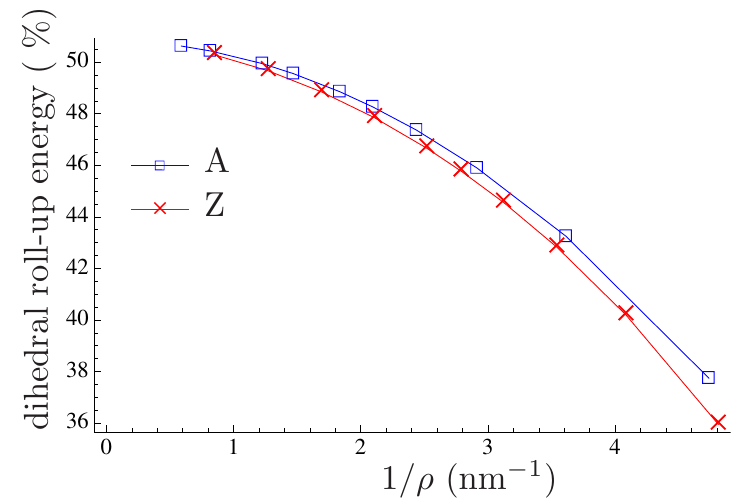}
\caption{Dihedral contribution to the roll-up energy \emph{vs.} natural curvature.}
\label{linearity}
\end{minipage}%
\hspace{10mm}%
\begin{minipage}[c]{.40\textwidth}
\includegraphics[scale=0.7]{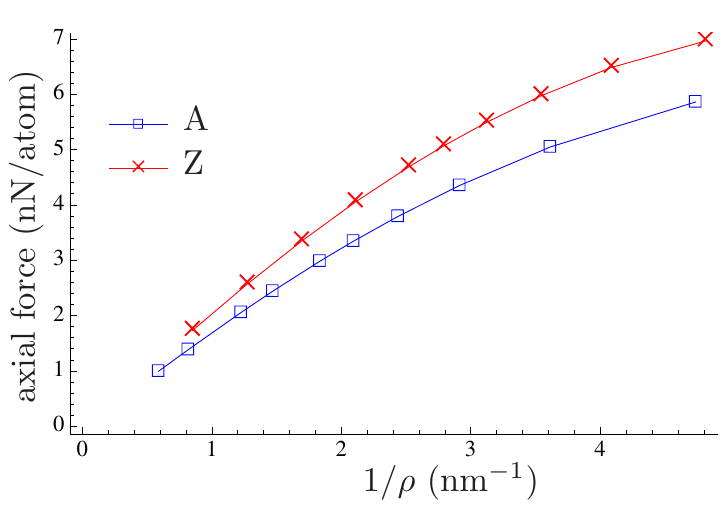}
\caption{ axial force  to be applied to each rim atom to have an energy increment equal to the roll-up energy \emph{vs} curvature.}
\label{prestrax}
\end{minipage}
\end{figure}

The dihedral contribution to the roll-up energy is depicted \emph{vs.} natural curvature in Fig. \ref{linearity}.
%
%
We see that this contribution decreases when curvature increases, a fact indicating that changes in bond angles have a prevailing role for small radii: the lower the curvature the smaller (larger) the bond-angle (dihedral-angle) roll-up energy. For large diameters, our numerical results are in complete agreement with those of \cite{Lu_2009}.
The numerical information we used to construct Figs. \ref{energyvsd2} and \ref{linearity} is collected in Table \ref{tab:energy}, Appendix D.

To better appreciate the relevance of self-stress in CNT mechanics, we have computed what axial traction uniformly applied along the rim would induce in a CNT an additional energy storage equal to its roll-up energy; this we did by exploiting again the  solution of the axial traction problem formulated and solved in Appendix C. Our results are plotted in Fig. \ref{prestrax}; we see that $f$, the axial traction to be applied to each rim atom in order to have an energy increment equal to the roll-up energy, is a monotonically increasing function of natural curvature (e.g., it is 2.05 nN/atom for a $(12,12)$-CNT and it goes up to 5.86 nN/atom for a $(3,3)$-CNT; for more information, see  Table \ref{tab:pestrax} in Appendix D). For $F$ the total traction applied to the rim, $F/2\pi\rho$, the load per unit length, should be compared with the ultimate tensile load of single-wall CNTs. Due to their small size, testing individual CNTs is very difficult. Nevertheless, according to \cite{Demczyk2002,Troiani2003,Ding2007}, their ultimate tensile load is of the same order of magnitude as  graphene's, that is, about $\simeq40$ nN/nm (see the last column of Table \ref{tab:pestrax}, where it is shown that, for achiral CNTs whose radius ranges from 0.20778 nm to 1.17601 nm, the load inducing the same energy as the roll-up energy approximately ranges from 66\% to 11\% of the ultimate load).

\subsection{The self-stress state of chiral CNTs}\label{chiral}

We carried out a numerical minimization of the potential \eqref{Vchi}, with the chiral order-parameter string $\widehat\csib^{\,c}\!(\qb)$ expressed in  terms of the same  Lagrangian coordinates as in \cite{Popov2004}.
As expected, we found that the self-energy of a chiral CNT depends on its size more or less like it does in the achiral case, with approximately the same roll-up energy difference with respect to the thin-plate model, and approximately the same dihedral contribution to the roll-up energy (cf. Tab. \ref{tab:chiral} and  Tab. \ref{tab:chiralen} in Appendix D).

\begin{figure}[H]
\centering
\qquad \hspace{1.cm}
\includegraphics[width=0.45\textwidth]{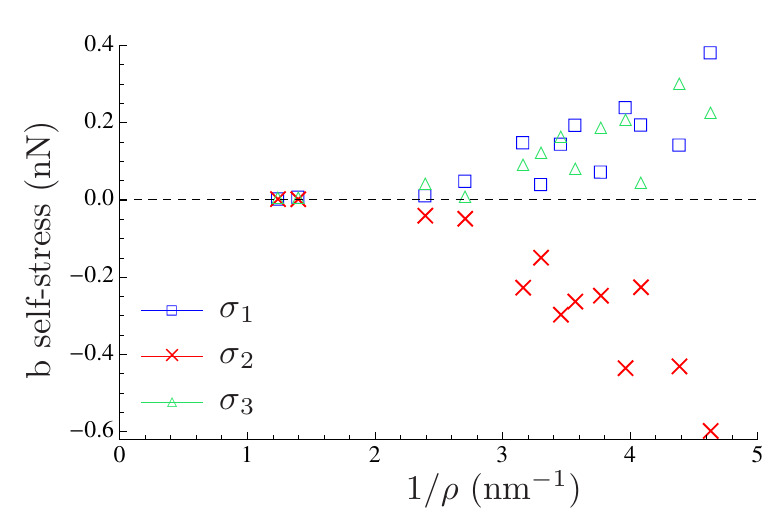}
\\[20pt]
\includegraphics[width=0.45\textwidth]{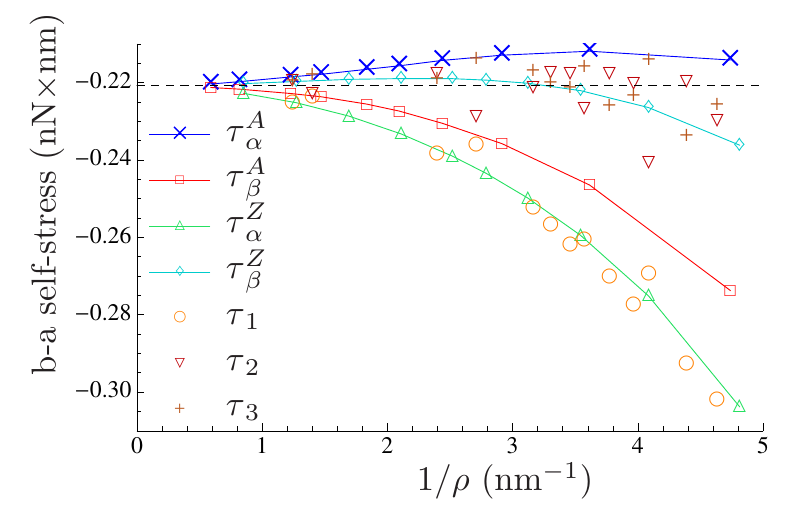}
\qquad
\includegraphics[width=0.45\textwidth]{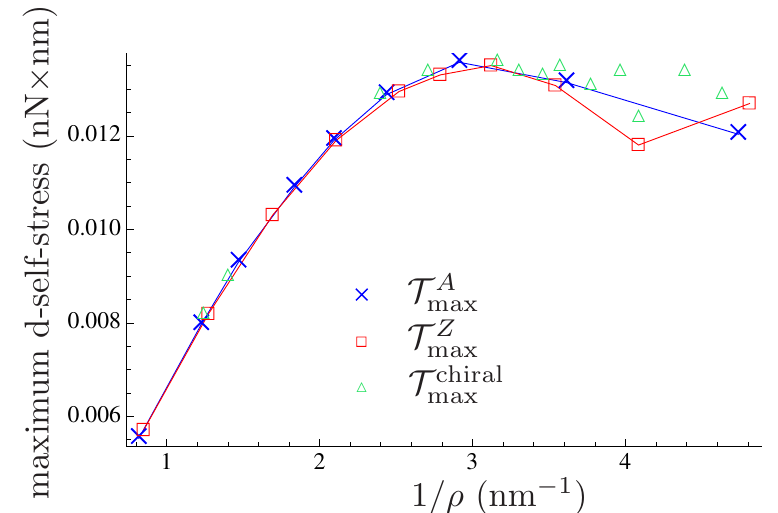}
\caption{(top left) Nanostresses vs. natural curvature in chiral CNTs: bond-length nanostress (top right), bond-angle nanostress (bottom left), and maximum dihedral-angle nanostress (bottom right); the blue and red lines in the bottom right figure represent the envelop of maximum dihedral nanostresses in achiral CNTs.} \label{nanostresschi}
\end{figure}
The main qualitative difference between chiral and achiral CNTs regards the bond-length nanostresses, which cannot be all null in the chiral case (recall the developments in Section \ref{Equil2}). However, as shown in Fig.\ref{nanostresschi} (top right), these nanostresses are rather small, in the range $[-0.6,0.6]$ nN. For a comparison, to obtain a value of $0.6$ nN of the bond-length stress in graphene, a uniform traction load of $3.23$ nN/nm ($2.44$ nN/nm), that is, less than 1/10 of the failure load, should be applied along the direction $\hb^Z$ ($\hb^A$). As to the bond-angle and dihedral-angle nanostresses, their values are in the same range as in the achiral case (cf Fig.\,\ref{nanostresschi} (bottom) with Fig.\,\ref{nanostressA}) (for further information, see Table \ref{tab:chiral} in Appendix D).

\section{Conclusions}

We have proposed a geometrically and physically nonlinear discrete mechanical model of single-wall carbon nanotubes, both achiral and chiral, and flat graphene strips. We have shown that \emph{all these carbon allotropes are stressed
in their equilibrium configurations prior to the application of external loads}. We have termed such configurations \emph{natural}, the relative stress distributions  \emph{self-stress states}, and the relative stored energy \emph{self-energy}. We have shown that self-stresses and self-energies are by no means negligible with respect to load-related energies. Two other types of energies we have computed and compared with self-energy are: the \emph{folding energy}, that is, the fictitious self-energy assigned to a CNT by the use of a formula borrowed from the cylindrical bending of thin elastic plates;  and the \emph{roll-up}  \emph{energy}, that is, the difference in energy per atom of a CNT  and its `parent' FGS, in their respective natural configurations.

Atomic interactions have been specified by a suitable form of the 2nd-generation Brenner potential, in terms of convenient \emph{order-parameter strings} consisting of bond lengths, bond angles, and dihedral angles. Governing equations have been written  in terms of \textit{nanostresses}, i.e., force-like objects one by one work-conjugated to the kinematic descriptors just listed: precisely, a self-stress state is nothing but a string of equilibrium nanostresses work-conjugated to
(i) changes in length of  atomic bonds; (ii) changes in angle between two adjacent bonds; (iii) changes in dihedral angles. Nonlinear constitutive equations for individual nanostresses in terms of these changes have been written. 
Equilibrium solutions for armchair and zigzag CNTs whose radii ranged from very large to very small have been obtained numerically, with minimal computational effort, by the use of slightly adjusted but otherwise standard computational tools; moreover, a body of qualitative and quantitative results have been presented in terms of self-energy and self-stresses for a number of chiral CNTs.

In our opinion, our major achievement consists in characterizing the natural equilibria of CNTs and FGSs and, in particular, in  identifying and evaluating quantitatively the various individual sources of self-stress states in such carbon allotropes. We surmise that our predictions may serve well as benchmarks for whatever MD code implementing REBO potentials.

\addcontentsline{toc}{section}{Acknowledgements}
\section*{Acknowledgements}
  NP is supported by the European Research
  Council (ERC StG Ideas 2011 BIHSNAM n. 279985 on ``Bio-Inspired hierarchical super-nanomaterials'', ERC PoC 2013-1 REPLICA2 n. 619448 on ``Large-area replication of biological antiadhesive nanosurfaces'', ERC PoC 2013-2 KNOTOUGH n. 632277 on `Super-tough knotted fibres''), by the European Commission under the Graphene Flagship (WP10 ``Nanocomposites'', n. 604391) and by the Provincia Autonoma di Trento (``Graphene Nanocomposites'', n. S116/2012-242637 and reg. delib. n. 2266). AF  acknowledges the Italian INdAM-GNFM (Istituto Nazionale di Alta Matematica -- Gruppo Nazionale di Fisica Matematica), through ``Progetto Giovani 2014 --  Mathematical models for complex nano- and bio-materials''. We all acknowledge interesting discussions with Dr.~Seunghwa Ryu about comparing our predictions with those of congruent MD simulations.
 \vfill
\newpage

\addcontentsline{toc}{section}{Appendix A: Details about the computation of dihedral angles}
\section*{Appendix A: Details about the computation of dihedral angles}

\subsection*{A1. Armchair CNTs}
With reference to Fig. \ref{cell}, and for $\cb_1,\cb_2,\cb_3$  the basis vectors of the Cartesian frame there shown, let us introduce the following unit vectors:
\begin{equation}
\begin{aligned}
&\ab:={\rm vers}\, \overrightarrow{HB_2}=-\cos\varphi^A\cb_1+\sin\varphi^A\cb_3,\\
&\ab_1:={\rm vers}\, \overrightarrow{H'A_2}=\cos\varphi^A\cb_1+\sin\varphi^A\cb_3,\\
&\ab_2:={\rm vers}\, \overrightarrow{B_2A_2''}=-\cos2\varphi^A\cb_1+\sin2\varphi^A\cb_3,\\
& \bb:={\rm vers}\, \overrightarrow{A_1B_1}=\cb_1,\\
& \cb:={\rm vers}\, \overrightarrow{A_1B_2}=\cos\frac{\alpha}{2}\ab+\sin\frac{\alpha}{2}\cb_2,\\
& \db:={\rm vers}\, \overrightarrow{A_1B_2'}=\cos\frac{\alpha}{2}\ab-\sin\frac{\alpha}{2}\cb_2,\\
& \db_1:={\rm vers}\, \overrightarrow{B_1A_2'}=\cos\frac{\alpha}{2}\ab_1-\sin\frac{\alpha}{2}\cb_2.
\end{aligned}
\end{equation}
In terms of these unit vectors, the cosines of the dihedral angles $\Theta_1^A,\Theta_2^A,\Theta_3^A$ read:
\begin{equation}
\begin{aligned}
&\cos\Theta_1^A=\frac{\bb\times\cb}{|\bb\times\cb|}\cdot\frac{\db_1\times\bb}{|\db_1\times\bb|}=\frac{1}{\sin^2\beta^A}\big(\sin^2\frac{\alpha}{2}-\cos^2\frac{\alpha}{2}\sin^2\varphi^A\big)\,,\\
& \cos\Theta_2^A=\frac{\cb\times\ab_2}{|\cb\times\ab_2|}\cdot\frac{\bb\times\cb}{|\bb\times\cb|}=\frac{1}{\sin^2\beta^A}\left( \cos^2\varphi^A\sin^2\frac{\alpha}{2}-\sin^2\varphi^A \right)\,,\\
& \cos\Theta_3^A=\frac{\cb\times\ab_2}{|\cb\times\ab_2|}\cdot\frac{\db\times\cb}{|\db\times\cb|}=-\frac{1}{\sin\beta^A}\,\sin\frac{\alpha}{2}\cos{\varphi^A}\,.
\end{aligned}
\end{equation}
Together with \eqref{geomcompA}, these relations, which are equivalent to the first three of \eqref{gamma1A}, permit to compute the first derivatives of $\beta^A$ and $\Theta_i^A \;(i=1,2,3)$ with respect to $\alpha$ :
\begin{equation}\label{betader}
\begin{aligned}
\beta^A,_\alpha &=-\frac{\sin\frac{\alpha}{2}\cos\varphi^A}{2\,\sin\beta^A}\,,
\\
\Theta_1^A,_\alpha &= -\frac{\sin \frac{\alpha }{2} \sin\varphi^A+2 \beta^A,_\alpha \cos\beta^A \sin
   \frac{\Theta_1^A}{2}}{\sin\beta^A\cos\frac{\Theta_1^A}{2} }\,,\\ \Theta_2^A,_\alpha &=-\frac{\beta^A,_\alpha \cos\beta^A \sin
      \Theta_2^A}{\sin\beta^A\cos\Theta_2^A}\,, \\ \Theta_3^A,_\alpha &=2\,\Theta_2^A,_\alpha \,.
\end{aligned}
\end{equation}

\subsection*{A2. Zigzag CNTs}
With reference to Fig. \ref{zigzag}, we introduce the unit vectors
\begin{equation}
\begin{aligned}
&\ab:={\rm vers}\,\overrightarrow{A_1H}=\cos\varphi^Z\cb_2+\sin\varphi^Z\cb_3,\\
&\ab_1:={\rm vers}\,\overrightarrow{HA_3}=\cos 2\varphi^Z\cb_2+\sin 2\varphi^Z\cb_3,\\
&\bb:={\rm vers}\,\overrightarrow{A_1B_2'}=-\sin\beta^Z\cb_2+\cos\beta^Z\cb_1,\\
&\bb_1:={\rm vers}\,\overrightarrow{B_1A_2'}=-\sin\beta^Z\cb_2-\cos\beta^Z\cb_1,\\
&\db:={\rm vers}\,\overrightarrow{A_1B_2}=\sin\beta^Z\ab+\cos\beta^Z\cb_1,\\
&\db_1:={\rm vers}\,\overrightarrow{B_1A_2}=\sin\beta^Z\ab-\cos\beta^Z\cb_1,\\
&\db_2:={\rm vers}\,\overrightarrow{A_2B_3}=\sin\beta^Z\ab_1+\cos\beta^Z\cb_1.\\
\end{aligned}
\end{equation}
For the dihedral angles, we find :
\begin{equation}
\begin{aligned}
&\cos\Theta_1^Z=\frac{\db_1\times\cb_1}{\sin\beta^Z}\cdot\frac{\cb\times\bb}{\sin\beta^Z}=\cos\varphi^Z\,,\\
&\cos\Theta_2^Z=\frac{\db_1\times\db_2}{\sin\alpha}\cdot\frac{\bb_1\times\db_1}{\sin\alpha}=\frac{1}{\sin^2\alpha}\,(\sin^2\alpha-2\sin^2\beta^Z\sin^2\varphi^Z)\,,\\
& \cos\Theta_4^Z=\frac{\db_1\times\db_2}{\sin\beta^Z}\cdot\frac{\cb_1\times\db_1}{\sin\beta^Z}=(1+\cos\varphi^Z)\cos\beta^Z\,,
\end{aligned}
\end{equation}
a set of relations equivalent to the first, second, and fourth, of \eqref{gamma1Z}. With these and \eqref{geomcompZ}, we find:
\begin{equation}\label{betaderZ}
\begin{aligned}
\beta^Z,_\alpha &=\frac{\cos\frac{\alpha}{2}}{2\,\cos\beta^Z\cos\frac{\varphi^Z}{2}}\,,
\\
\Theta_{1,\alpha}^Z &=0\,, \\ \Theta_2^Z,_\alpha &= \frac{\beta^Z,_\alpha \cos\beta^Z \sin
   \varphi^Z- \cos\alpha\sin\Theta_2^Z}{\sin\alpha\cos\Theta_2^Z}\,,  \\ \Theta_4^Z,_\alpha &=2\, \Theta_2^Z,_\alpha \,.
   \end{aligned}
\end{equation}

%
%


\addcontentsline{toc}{section}{Appendix B: 2nd-generation REBO potentials for hexagonal lattices}
\section*{Appendix B:\\[3pt]2nd-generation REBO potentials for hexagonal lattices}

\subsection*{B1. General form.} As anticipated in Section \ref{natequi}, the 2nd-generation REBO potentials developed for hydrocarbons by Brenner \emph{et al.} in \cite{Brenner_2002} accommodate  up third-nearest-neighbor interactions, through a bond-order function depending also on dihedral angles. In general, given a substance or a group of substances in the hydrocarbon family, the appropriate potential is tailored by fitting the parameters to the available experimental data and \emph{ab initio} calculations; the behavior of electron clouds is not accounted for explicitly, and  quantum effects are ignored. In spite of these limitations, the predictions obtained with the use of  REBO potentials, when compared with those obtained by \emph{ab initio} or TB methods, have been always found accurate qualitatively, and sometimes even quantitatively. In fact, REBO potentials do incorporate much of the physics and chemistry involved in covalent bonding, as well as Coulomb interactions and many-body effects; if necessary, they also can accommodate bond-breaking and bond-formation.

The  binding energy $V$ of an atomic aggregate is written as a sum over nearest neighbors:
\begin{equation}\label{V}
V=\sum_I\sum_{J<I} V_{IJ}\,;
\end{equation}
 the interatomic potential $V_{IJ}$ is given by the construct
\begin{equation}\label{Vij}
V_{IJ}=V_R(r_{IJ})+b_{IJ}V_A(r_{IJ}),
\end{equation}
where the individual effects of the \emph{repulsion} and \emph{attraction functions} $V_R(r_{IJ})$ and $V_A(r_{IJ})$, which model pair-wise interactions of  atoms $I$ and $J$ depending on their distance $r_{IJ}$, are modulated by the \emph{bond-order function} $b_{IJ}$. The repulsion and attraction functions have the following forms:
\begin{equation}\label{VA}
\begin{aligned}
V_A(r)&=-f^C(r)\sum_{n=1}^{3}B_n e^{-\beta_n r}\,,\\
V_R(p)&=f^C(r)\left( 1 + \frac{Q}{r} \right) A e^{-\alpha r}\,,
\end{aligned}
\end{equation}
where $f^C(r)$ is a \emph{cutoff function} limiting the range of covalent interactions, and where $Q$, $A$, $B_n$, $\alpha$, and $\beta$, are parameters to be chosen fit to a material-specific dataset. The remaining ingredient in \eqref{Vij} is the \emph{bond-order function}:
\begin{equation}\label{bij}
b_{IJ}=\frac{1}{2}(b_{IJ}^{\sigma-\pi}+b_{JI}^{\sigma-\pi})+b_{IJ}^\pi\,,
\end{equation}
where apexes $\sigma$ and $\pi$ refer to two types of bonds: the strong covalent $\sigma$-bonds between atoms in one and the same given plane, and the $\pi$-bonds responsible for interlayer interactions, which are perpendicular to the plane of $\sigma$-bonds. We now describe functions $b_{IJ}^{\sigma-\pi}$ and $b_{IJ}^\pi$.

The role of function $b_{IJ}^{\sigma-\pi}$ is to account for the local coordination of, and the bond angles relative to, atoms $I$ and $J$; its form is:
\begin{equation}\label{G}
b_{IJ}^{\sigma-\pi}=\left(1+\sum_{K\neq I,J} f_{IK}^C(r_{IK})G(\cos\theta_{IJK}) \,e^{\lambda_{IJK}}+P_{IJ}(N_I^C,N_I^H)  \right)^{-1/2}\,.
\end{equation}
Here, for each fixed pair of indices $(I,J)$, (a) the cutoff function $f_{IK}^C$ limits the interactions of atom $I$ to those with its nearest neighbors; (b) $\lambda_{IJK}$ is a string of parameters designed to prevent attraction in some specific situations; (c) function $P_{IJ}$ depends on $N_I^C$ and $N_I^H$, the numbers of $C$ and $H$ atoms that are nearest neighbors of atom $I$; it is meant to adjust the bond-order function 
according to the environment of the C atoms in one or another molecule; (d) for solid-state carbon, the values of  both the string $\lambda_{IJK}$ and the function $P_{IJ}$ are taken null; (e)  function $G$ modulates the contribution of each nearest neighbour of atom $I$ in terms of the cosine of the angle between the $IJ$ and $IK$ bonds; its analytic form is given by three sixth-order polynomial splines. 

Function $b_{IJ}^\pi$ is given a split representation:
\begin{equation}
b_{IJ}^\pi=\Pi_{IJ}^{RC}+b_{IJ}^{DH},
\end{equation}
where the first addendum $\Pi_{IJ}^{RC}$ depends on whether the bond between atoms $I$ and $J$ has a radical character and on whether it is part of a conjugated system, while the second addendum $b_{IJ}^{DH}$ depends on dihedral angles and has the following form:
\begin{equation}
b_{IJ}^{DH}=T_{IJ}(N_I^t,N_J^t,N_{IJ}^{\rm conj})\left(\sum_{K(\neq I,J)}\sum_{K(\neq I,J)}\big( 1-\cos^2\Theta_{IJKL} \big)f_{IK}^C(r_{IK})f_{JL}^C(r_{JL})  \right)\,,
\end{equation}
where function $T_{IJ}$ is a tricubic spline depending on $N_I^t=N_I^C+N_I^H$, $N_J^t$, and $N_{IJ}^{\rm conj}$, a function  of local conjugation. 



\subsection*{B.2 The form used in this paper} We write here the expressions we use  for the functions $V_A$, $V_R$ and $b_{IJ}$ defined in \eqref{VA} and \eqref{bij}; we also record the form of their first derivatives, because they enter equations \eqref{nanostress} and \eqref{nnstresschi}.

The attractive and repulsive part of the potential, and their first derivatives are:
\begin{equation}\label{VAR}
\begin{aligned}
& V_A(r)=-\sum_{n=1}^{3}B_n e^{-\beta_n r}\,,\quad
V'_A(r)=\sum_{n=1}^{3}\beta_n B_n e^{-\beta_n r}\,,\\
& V_R(r)=\left( 1 + \frac{Q}{r} \right) A e^{-\alpha r}\,,\quad
V'_R(r)=-\frac{Q}{r^2} A e^{-\alpha r} - \alpha \left( 1 + \frac{Q}{r} \right) A e^{-\alpha r}\,.
\end{aligned}
\end{equation}

%
%
The bond order function $b_{IJ}$ \eqref{bij} specializes to $b_a$ and $b_b$,  respectively, for $a$- and $b$-type bonds 
in achiral CNTs (cf. equations \eqref{Va:AZ}), and it specializes to $b_i$ for the typical bond in chiral CNTs (cf. the second of \eqref{Vchi}):
\begin{equation}\label{bab}
\begin{aligned}
& b_{a}=\big( 1+ 2 G(\beta ) \big)^{-\frac{1}{2}} + 2 T ( 1 - \cos^2 \Theta_1 )\,,\\
& b_{b}=\big( 1+ G(\alpha)+G(\beta ) \big)^{-\frac{1}{2}} +
T\big( 2( 1 - \cos^2 \Theta_2 )+( 1 - \cos^2 \Theta_3 )+( 1 - \cos^2 \Theta_4 )\big)\,,\\
&
b_{i}=\big( 1+ G(\theta_{i+1})+G(\theta_{i+2}) \big)^{-\frac{1}{2}} +
T\big( 2( 1 - \cos^2 \Theta_{i1} )+( 1 - \cos^2 \Theta_{i2} )+( 1 - \cos^2 \Theta_{i3} )\big)
\,.
\end{aligned}
\end{equation}
%
The following derivatives of  bond-order functions are found in equation \eqref{nanostress}:
\begin{align}
& b_{a},_\beta = -\big( 1+ 2 G(\beta ) \big)^{-\frac{3}{2}} G'(\beta)\,,\label{bab}\\
& b_{a},_{\Theta_1} = 4 T \cos\Theta_1\sin\Theta_1\,,\label{bag}\\
& b_{b},_\alpha = -\frac{1}{2}\big( 1+ G(\alpha)+ G(\beta ) \big)^{-\frac{3}{2}} G'(\alpha)\,,\label{bba}\\
& b_{b},_\beta = -\frac{1}{2}\big( 1+ G(\alpha)+ G(\beta ) \big)^{-\frac{3}{2}} G'(\beta)\,,\label{bbb}\\
& b_{b},_{\Theta_2} = 4 T \cos\Theta_2\sin\Theta_2\,,\label{bbg2}\\
& b_{b},_{\Theta_3} = 2 T \cos\Theta_3\sin\Theta_3\,,\label{bbg3}\\
& b_{b},_{\Theta_4} = 2 T \cos\Theta_4\sin\Theta_4\,.\label{bbg4};
\end{align}
the derivatives found in equation \eqref{nnstresschi}  are:
\begin{align}
& (b_{i+1}),_{\theta i}=-\frac{1}{2}\big( 1+ G(\theta_i)+ G(\theta_{i+2}) \big)^{-\frac{3}{2}} G'(\theta_i)\,,\\
& b_{i},_{\Theta_{i1}} = 4 T \cos\Theta_{i1}\sin\Theta_{i1}\,,\qquad
b_{i},_{\Theta_{ij}} = 2 T \cos\Theta_{ij}\sin\Theta_{ij}\,,\qquad i=1,2,3,\quad j=2,3\\
\end{align}
(subscripts should be taken \emph{modulo} 3).

In equation \eqref{G}, the angular-contribution function $G$ is:
\begin{equation}\label{Gfun}
G(\theta)=\left\{
\begin{array}{ll}
  G_1(\theta)\,, & 0\leq\theta<0.6082\pi\\
  G_2(\theta)\,, & 0.6082\pi\leq\theta<\frac{2\pi}{3} \\
  G_3(\theta)\,, & \frac{2\pi}{3}\leq\theta\leq\pi
\end{array}
\right.\,, \quad G_j(\theta)=\sum_{i=0}^{5}d_{ji}\,(\cos\theta)^i\,,\ j=1,2,3\,,
\end{equation}
%
whence
\begin{equation}\label{Giprime}
G'_j(\theta)=-\sin\theta\left(\sum_{i=1}^{5} i\, d_{ji}\,(\cos\theta)^{i-1}\right)\,,\ j=1,2,3\,.
\end{equation}
The polynomial coefficients $d_{ji}$ are computed following \cite{Brenner_2002}; they are reported in the following table:
\begin{center}
\begin{tabular}{cc|cccccc}
$d_{ji}$ & \ & \ & \ & \ \quad \qquad $i$ \\
\ & \ & 0 & 1 & 2 & 3 & 4 & 5 \\
\hline
\ & 1 & 0.37545 & 1.40678 & 2.25438 & 2.03128 & 1.42971 & 0.50240\\
$j$ & 2 & 0.70728 & 5.67744 & 24.09702 & 57.59183 & 71.88287 & 36.27886\\
\ & 3 & -0.64440 & -6.20800 & -20.05900 & -30.22800 & -21.72400 & -5.99040\\
\end{tabular}
\end{center}

The parameters of the binding energy $V$, the same as in \cite{Brenner_2002,Stuart2000}, are:

\begin{center}
\begin{tabular}{lll}
$B_1$ = 12388.79197798 eV\,, & $\beta_1$ = 47.204523127 nm$^{-1}$\,, & $Q$ = 0.03134602960833 nm\,, \\
$B_2$ = 17.56740646509 eV\,, & $\beta_2$ = 14.332132499 nm$^{-1}$\,, & $A$ = 10953.544162170 eV\,, \\
$B_3$ = 30.71493208065 eV\,, & $\beta_3$ = 13.826912506 nm$^{-1}$\,, & $\alpha$ = 47.465390606595 nm$^{-1}$\,, \\
$T$ = $-$0.004048375\,. \\
\end{tabular}
\end{center}

%

\addcontentsline{toc}{section}{Appendix C: The traction problem}
\section*{Appendix C: The traction problem}\label{traction}
We here derive the explicit form of the balance equations for the case when a pure-traction load is applied to an achiral CNT.  We made use of the solution to this problem in Sect. \ref{SSS}, when we compared the folding energy of CNTs with the energy stored in such a traction problem.

With reference to \eqref{totpot}, for $F$ be the magnitude of the axial traction, the load potential takes the following form:
\begin{equation}
\fb\cdot\widehat{\db}(\qb)=F\delta\widehat{\lambda}(a,b,\alpha),
\end{equation}
where $\delta\widehat{\lambda}(a,b,\alpha)$ is the load-induced change in length of the CNT under study. It follows that the balance equations are:

\begin{equation}\label{finalitrA}
\begin{aligned}
&\sigma_a= F\frac{\widehat{\lambda},_{a}}{n_1\,n_2}\,,\\
&\sigma_b= F\frac{\widehat{\lambda},_{b}}{2\,n_1\,n_2}\,,\\
&\tau_\alpha+2\,\beta,_\alpha\, \tau_\beta +\sum_{i=1}^4 \,\Theta_i,_\alpha\,\Tc_i= F\frac{\widehat{\lambda},_{\alpha}}{2\,n_1\,n_2}\,.
\end{aligned}
\end{equation}
The mappings $(a,b,\alpha)\mapsto\widehat{\lambda}^A(a,b,\alpha)$ and $(a,b,\alpha)\mapsto\widehat{\lambda}^Z(a,b,\alpha)$ are here defined with the use of equations \eqref{L2A} and \eqref{L1Z}), respectively; hence, we have that:
\begin{equation}\label{finalitrB}
\left[\begin{array}{c}
\widehat{\lambda^A},_{a}(a,b,\alpha) \\
\widehat{\lambda^A},_{b}(a,b,\alpha)  \\
\widehat{\lambda^A},_{\alpha}(a,b,\alpha)\\
\end{array}\right]\,=
\left[\begin{array}{c}
0 \\
2n_2\sin\frac{\alpha}{2} \\
n_2b\cos\frac{\alpha}{2}\\
\end{array}\right]\,, \quad
\left[\begin{array}{c}
\widehat{\lambda^Z},_{a}(a,b,\alpha) \\
\widehat{\lambda^Z},_{b}(a,b,\alpha)  \\
\widehat{\lambda^Z},_{\alpha}(a,b,\alpha)\\
\end{array}\right]\,=
\left[\begin{array}{c}
n_1 \\
-n_1\cos\beta^Z \\
bn_1\beta^Z,_{\alpha}\sin\beta^Z\\
\end{array}\right]\,.
\end{equation}

Equations \eqref{finalitrA}-\eqref{finalitrB} can be so specialized as to hold in the case of a FGS subject to a uniform traction load along the armchair or the zigzag direction: it is enough to take  $\varphi=0$ and, consequently, $\alpha+2\beta=2\pi$, $\Theta_i=0$, $i=1,\ldots,4$.

\addcontentsline{toc}{section}{Appendix D: Computational results}
\section*{Appendix D: Computational results}\label{AC}
In this Appendix we collect a number of tables summarizing the results of our computations.
Numerical values for the natural geometric parameters are shown in Tables \ref{tab:geometry} and \ref{tab:dh:angles}. Table \ref{tab:geometry} also shows: (i) the percent difference of bond lengths $a$ and $b$ with respect to $r_0=0.14204$\,nm, the C-C distance in graphene computed according to the potential chosen in this study; (ii) the percent differences between the natural and nominal values of the bond angles $\alpha$ and $\beta$ ($\alpha_0$ and $\beta_0$ have been computed by substituting $\alpha_0^A=2\pi/3$ in \eqref{betafunA} (A case) and $\beta_0^Z=2\pi/3$ in \eqref{geomcompZ}, and then solving for $\alpha_0^Z$ (Z case)). 

\begin{table}[h!]
\begin{center}
\caption{Natural geometry of achiral CNTs.}
\label{tab:geometry}
\vskip 1pt
\begin{tabular}{cccccccccccc}
$(n,m)$ & $\rho$ & $\frac{\rho-\rho_0}{\rho_0}$  &  $\frac{\lambda-\lambda_0}{\lambda_0}$ & $a$ &  $\frac{a-r_0}{r_0}$ & $b$ &  $\frac{b-r_0}{r_0}$ & $\alpha$ &  $\frac{\alpha-\alpha_0}{\alpha_0}$ & $\beta$ &  $\frac{\beta-\beta_0}{\beta_0}$ \\[5pt]
\ & (nm) & (\%)  &  (\%) & (nm) &  (\%) & (nm) &  (\%) & (degrees) &  (\%) & (degrees) &  (\%) \\[3pt]
\hline
(3,3) & 0.21111 & 3.76 & 0.94 & 0.14425 & 1.56 & 0.14428 & 1.58 & 118.76 & -1.03 & 116.17 & 0.44 \\
(4,4) & 0.27666 & 1.99 & 0.72 & 0.14331 & 0.90 & 0.14336 & 0.93 & 119.59 & -0.34 & 117.70 & 0.16 \\
(5,5) & 0.34313 & 1.19 & 0.57 & 0.14289 & 0.60 & 0.14290 & 0.60 & 119.93 & -0.06 & 118.43 & 0.03 \\
(6,6) & 0.41008 & 0.78 & 0.45 & 0.14265 & 0.43 & 0.14264 & 0.42 & 120.06 & 0.05 & 118.85 & -0.03 \\
(7,7) & 0.47733 & 0.55 & 0.37 & 0.14250 & 0.33 & 0.14248 & 0.31 & 120.11 & 0.09 & 119.12 & -0.04 \\
(8,8) & 0.54474 & 0.40 & 0.30 & 0.14240 & 0.25 & 0.14238 & 0.24 & 120.12 & 0.10 & 119.31 & -0.05 \\
(10,10) & 0.67985 & 0.24 & 0.21 & 0.14228 & 0.17 & 0.14225 & 0.15 & 120.11 & 0.09 & 119.54 & -0.04 \\
(12,12) & 0.81516 & 0.16 & 0.15 & 0.14221 & 0.12 & 0.14219 & 0.11 & 120.09 & 0.07 & 119.67 & -0.04 \\
(18,18) & 1.22158 & 0.07 & 0.07 & 0.14212 & 0.05 & 0.14210 & 0.05 & 120.05 & 0.04 & 119.85 & -0.02 \\
(25,25) & 1.69606 & 0.03 & 0.04 & 0.14208 & 0.03 & 0.14207 & 0.02 & 120.03 & 0.02 & 119.92 & -0.01 \\
\hline
(5,0) & 0.20778 & 6.13 & -1.03 & 0.14358 & 1.08 & 0.14497 & 2.06 & 114.80 & 10.75 & 117.65 & -1.96 \\
(6,0) & 0.24472 & 4.17 & -0.58 & 0.14305 & 0.71 & 0.14414 & 1.48 & 116.18 & 6.13 & 118.50 & -1.25 \\
(7,0) & 0.28227 & 2.99 & -0.32 & 0.14274 & 0.49 & 0.14363 & 1.12 & 117.01 & 3.93 & 119.00 & -0.83 \\
(8,0) & 0.32022 & 2.23 & -0.17 & 0.14254 & 0.36 & 0.14329 & 0.88 & 117.57 & 2.71 & 119.31 & -0.57 \\
(9,0) & 0.35845 & 1.72 & -0.08 & 0.14241 & 0.27 & 0.14305 & 0.71 & 117.97 & 1.95 & 119.51 & -0.40 \\
(10,0) & 0.39686 & 1.36 & -0.02 & 0.14233 & 0.20 & 0.14287 & 0.59 & 118.27 & 1.46 & 119.65 & -0.29 \\
(12,0) & 0.47409 & 0.90 & 0.03 & 0.14222 & 0.13 & 0.14264 & 0.42 & 118.69 & 0.88 & 119.81 & -0.16 \\
(15,0) & 0.59051 & 0.54 & 0.06 & 0.14214 & 0.07 & 0.14243 & 0.28 & 119.08 & 0.47 & 119.92 & -0.07 \\
(20,0) & 0.78532 & 0.28 & 0.06 & 0.14208 & 0.03 & 0.14227 & 0.16 & 119.43 & 0.21 & 119.98 & -0.02 \\
(30,0) & 1.17601 & 0.12 & 0.04 & 0.14205 & 0.01 & 0.14214 & 0.07 & 119.72 & 0.07 & 120.00 & 0.00 \\
\hline
\end{tabular}
\end{center}
\end{table}

\begin{table}[h!]
\begin{center}
\caption{Dihedral angles (degrees).}
\label{tab:dh:angles}
\vskip 3pt
\begin{tabular}{ccccc}
$(n,m)$ & $\Theta_1$ & $\Theta_2$  &  $\Theta_3$ & $\Theta_4$ \\[3pt]
\hline
(3,3) & 32.97 & 33.86 & 67.71 & 0 \\
(4,4) & 25.12 & 25.61 & 51.21 & 0 \\
(5,5) & 20.26 & 20.57 & 41.14 & 0 \\
(6,6) & 16.98 & 17.19 & 34.37 & 0 \\
(7,7) & 14.61 & 14.76 & 29.51 & 0 \\
(8,8) & 12.82 & 12.93 & 25.86 & 0 \\
(10,10) & 10.30 & 10.36 & 20.72 & 0 \\
(12,12) & 8.60 & 8.64 & 17.28 & 0 \\
(18,18) & 5.76 & 5.77 & 11.53 & 0 \\
(25,25) & 4.15 & 4.15 & 8.31 & 0 \\
\hline
\end{tabular}
\hspace{1cm}
\begin{tabular}{ccccc}
$(n,m)$ & $\Theta_1$ & $\Theta_2$  &  $\Theta_3$ & $\Theta_4$ \\[3pt]
\hline
(5,0) & 36.00 & 35.00 & 0 & 69.99 \\
(6,0) & 30.00 & 29.32 & 0 & 58.64 \\
(7,0) & 25.71 & 25.21 & 0 & 50.42 \\
(8,0) & 22.50 & 22.11 & 0 & 44.22 \\
(9,0) & 20.00 & 19.69 & 0 & 39.39 \\
(10,0) & 18.00 & 17.75 & 0 & 35.51 \\
(12,0) & 15.00 & 14.83 & 0 & 29.67 \\
(15,0) & 12.00 & 11.90 & 0 & 23.80 \\
(20,0) & 9.00 & 8.95 & 0 & 17.90 \\
(30,0) & 6.00 & 5.98 & 0 & 11.97 \\
\hline
\end{tabular}
\end{center}
\end{table}

\begin{table}[h]
\begin{center}
\caption{Geometry of CNTs using a 1st-generation  REBO potential (from \cite{Jiang_2003}).}
\label{Jiang:res}
\vskip 1pt
\begin{tabular}{ccccccc}
 $(n,m)$ &$\rho$ & $a$  & $b$  & $\alpha$ & $\beta$  &  \\
&  (nm) & (nm) & (nm)  & (degrees) & (degrees)  &  \\[3pt]
\hline
$(4,4)$ & 0.28036 & 0.14604 & 0.14548 & 120.45 & 117.31\\
$(5,5)$ & 0.34899 & 0.14568 & 0.14533 & 120.26 & 118.27\\
$(6,6)$ & 0.41782 & 0.14549 & 0.14525 & 120.17 & 118.80\\
$(12,12)$ & 0.83230 & 0.14517 & 0.14511 & 120.04 & 119.70\\
$(18,18)$ & 1.24750 & 0.14511 & 0.14509 & 120.02 & 119.87\\
\hline
$(5,0)$ & 0.20761 & 0.14542 & 0.14669 & 112.59 & 118.99\\
$(7,0)$ & 0.28552 & 0.14528 & 0.14586 & 116.28 & 119.40\\
$(10,0)$ & 0.40386 & 0.14518 & 0.14544 & 118.20 & 119.69\\
$(20,0)$ & 0.80180 & 0.14510 & 0.14516 & 119.55 & 119.92\\
$(30,0)$ & 1.20105 & 0.14508 & 0.14511 & 119.80 & 119.96 \\
\hline
\end{tabular}
\end{center}
\end{table}

\begin{table}[h!]
\begin{center}
\caption{Nanostresses associated to bond and dihedral angles (nN$\times$nm).}
\label{tab:nanostress}
\vskip 3pt
\begin{tabular}{ccccccc}
$(n,m)$ & $\tau_\alpha$ & $\tau_\beta$ & $\Tc_1$ & $\Tc_2$  &  $\Tc_3$ & $\Tc_4$ \\[3pt]
\hline
(3,3) & -0.2152 & -0.2730 & 0.0119 & 0.0120 & 0.0091 & 0 \\
(4,4) & -0.2126 & -0.2459 & 0.0104 & 0.0105 & 0.0131 & 0 \\
(5,5) & -0.2134 & -0.2354 & 0.0089 & 0.0090 & 0.0136 & 0 \\
(6,6) & -0.2146 & -0.2303 & 0.0077 & 0.0078 & 0.0129 & 0 \\
(7,7) & -0.2157 & -0.2274 & 0.0068 & 0.0068 & 0.0119 & 0 \\
(8,8) & -0.2166 & -0.2256 & 0.0060 & 0.0061 & 0.0109 & 0 \\
(10,10) & -0.2179 & -0.2237 & 0.0049 & 0.0050 & 0.0093 & 0 \\
(12,12) & -0.2187 & -0.2228 & 0.0042 & 0.0042 & 0.0080 & 0 \\
(18,18) & -0.2199 & -0.2217 & 0.0028 & 0.0028 & 0.0055 & 0 \\
(25,25) & -0.2204 & -0.2213 & 0.0020 & 0.0020 & 0.0040 & 0 \\
\hline
(5,0) & -0.3038 & -0.2362 & 0.0127 & 0.0119 & 0 & 0.0082 \\
(6,0) & -0.2752 & -0.2265 & 0.0118 & 0.0112 & 0 & 0.0116 \\
(7,0) & -0.2595 & -0.2221 & 0.0108 & 0.0103 & 0 & 0.0131 \\
(8,0) & -0.2499 & -0.2202 & 0.0098 & 0.0094 & 0 & 0.0135 \\
(9,0) & -0.2435 & -0.2194 & 0.0090 & 0.0086 & 0 & 0.0133 \\
(10,0) & -0.2391 & -0.2190 & 0.0082 & 0.0080 & 0 & 0.0129 \\
(12,0) & -0.2333 & -0.2190 & 0.0070 & 0.0068 & 0 & 0.0119 \\
(15,0) & -0.2287 & -0.2192 & 0.0057 & 0.0056 & 0 & 0.0103 \\
(20,0) & -0.2252 & -0.2197 & 0.0044 & 0.0043 & 0 & 0.0082 \\
(30,0) & -0.2228 & -0.2203 & 0.0029 & 0.0029 & 0 & 0.0057 \\
\hline
\end{tabular}
\end{center}
\end{table}

\begin{table}[h]
\begin{center}
\caption{Natural curvature, roll-up energy, percent difference of roll-up and `thin-plate' energy, dihedral contribution to the roll-up energy.}
\label{tab:energy}
\vskip 4pt
\begin{tabular}{ccccc}
\ & Natural & \   &  Difference & Dihedral \\
$(n,m)$ & curvature &  Roll-up  &  from $\frac{1}{2}\frac{D}{\rho^2}$ & contribution \\
\ & $1/\rho$ & energy  &  behavior & to the energy \\[3pt]
\ & (nm$^{-1}$) & (eV/atom)  &  (\%) & (\%) \\[3pt]
\hline
(3,3) & 4.7369 & 0.3814 & -7.5 & 37.7 \\
(4,4) & 3.6145 & 0.2256 & -6.0 & 43.2 \\
(5,5) & 2.9144 & 0.1489 & -4.6 & 45.9 \\
(6,6) & 2.4385 & 0.1054 & -3.6 & 47.4 \\
(7,7) & 2.0950 & 0.0784 & -2.8 & 48.3 \\
(8,8) & 1.8357 & 0.0605 & -2.3 & 48.8 \\
(10,10) & 1.4709 & 0.0391 & -1.5 & 49.5 \\
(12,12) & 1.2267 & 0.0273 & -1.1 & 49.9 \\
(18,18) & 0.8186 & 0.0122 & -0.5 & 50.4 \\
(25,25) & 0.5896 & 0.0064 & -0.3 & 50.6 \\
\hline
(5,0) & 4.8129 & 0.4193 & -1.5 & 35.9 \\
(6,0) & 4.0863 & 0.2982 & -2.8 & 40.2 \\
(7,0) & 3.5427 & 0.2235 & -3.1 & 42.8 \\
(8,0) & 3.1228 & 0.1738 & -3.0 & 44.6 \\
(9,0) & 2.7898 & 0.1390 & -2.8 & 45.8 \\
(10,0) & 2.5198 & 0.1137 & -2.5 & 46.7 \\
(12,0) & 2.1093 & 0.0800 & -2.1 & 47.8 \\
(15,0) & 1.6935 & 0.0518 & -1.6 & 48.8 \\
(20,0) & 1.2734 & 0.0294 & -1.0 & 49.7 \\
(30,0) & 0.8503 & 0.0132 & -0.5 & 50.3 \\
\hline
\end{tabular}
\end{center}
\end{table}

\begin{table}[h!]
\begin{center}
\caption{$f$ is the axial traction to be applied to each rim atom in order to have an energy increment equal to the roll-up energy; $F$ is the total axial traction;  $F/2\pi\rho$, the axial traction per unit rim length, should be compared with the ultimate load reported in the literature \cite{Demczyk2002}.}

\label{tab:pestrax}
\vskip 3pt
\begin{tabular}{cccccc}
\ & CNT  self-energy & roll-up energy  &  $f$ & $F$ & $F/2\pi\rho$\\[3pt]
$(n,m)$ & (eV/atom) & (eV/atom)  &  (nN/atom) & (nN) & (nN/nm)  \\[3pt]
\hline
(3,3)   & -7.0137 & 0.3814 & 5.8567 & 35.1402 &  26.4919  \\
(4,4)   & -7.1695 & 0.2256 & 5.0431 &  40.3447  &  23.2091 \\
(5,5)   & -7.2463 & 0.1489 &  4.3480  & 43.4800 & 20.1676  \\
(6,6)   & -7.2898 & 0.1054 &  3.7888  &  45.4661  &  17.6455 \\
(7,7)   & -7.3167 & 0.0784 & 3.3413  &  46.7778  & 15.5971 \\
(8,8)   & -7.3346 & 0.0605 & 2.9795  &   47.6715 & 13.9280 \\
(10,10) & -7.3560 & 0.0391 &  2.4367  &  48.7350 &  11.4090 \\
(12,12) & -7.3678 & 0.0273 &  2.0531 &  49.2746  &  9.6205 \\
(18,18) & -7.3829 & 0.0122 &  1.3810  &  49.7152 & 6.4772 \\
(25,25) & -7.3887 & 0.0064 &  0.9917 &  49.5842   & 4.6530    \\
\hline
(5,0) & -6.9758 & 0.4193 &  6.9570  &  34.7852  &  26.6452 \\
(6,0) & -7.0969 & 0.2982 &  6.4790  &  38.874 & 25.2821 \\
(7,0) & -7.1715 & 0.2236 &  5.9658  &  41.7607 &  23.5461\\
(8,0) & -7.2212 & 0.1739 &  5.4881  &  43.9049  &  21.8213\\
(9,0) & -7.2560 & 0.1391 &  5.0596 &  45.5366 &  20.2189\\
(10,0) & -7.2814 & 0.1137 &  4.6799 &  46.7993 &  18.7681 \\
(12,0) & -7.3151 & 0.0801 &  4.0480 &  48.5755&  16.3072 \\
(15,0) & -7.3432 & 0.0519 &  3.3415 &  50.1225  & 13.5090 \\
(20,0) & -7.3656 & 0.0295 &  2.5643&  51.2859  &  10.3937 \\
(30,0) & -7.3819 & 0.0132 &  1.7261 &  51.7824 &  7.0080 \\
\hline
\end{tabular}
\end{center}
\end{table}

\begin{table}[h!]
\begin{center}
\caption{Natural radius and nanostresses  in chiral CNTs.}
\label{tab:chiral}
\vskip 3pt
\begin{tabular}{cccccccccc}
\ & $\rho$ & $\sigma_a$ & $\sigma_b$  &  $\sigma_c$ & $\tau_a$ & $\tau_b$ & $\tau_c$ & max $\mathcal{T}$  \\[3pt]
$(n,m)$& (nN) & (nN) & (nN)  &  (nN) & (nN$\times$nm) & (nN$\times$nm) & (nN$\times$nm) & (nN$\times$nm)  \\[3pt]
\hline
(20,1) & 0.8054&  0.0004 & -0.0044 & 0.0040 & -0.2254 & -0.2197 & -0.2196 & 0.0082 \\
(11,10) & 0.7139 & 0.0045 & -0.0048 & 0.0003 & -0.2240 & -0.2231 & -0.2180 & 0.0090 \\
(10,1) & 0.4173 & 0.0083 & -0.0474 & 0.0391 & -0.2387 & -0.2180 & -0.2190 & 0.0129 \\
(6,5) & 0.3771 & 0.0462 & -0.0522 & 0.0060 & -0.2362 & -0.2290 & -0.2139 & 0.0134 \\
(6,3) & 0.3161 & 0.1459 & -0.2336 & 0.0877 & -0.2527 & -0.2216 & -0.2169 & 0.0136 \\
(7,1) & 0.3027 & 0.0374 & -0.1559 & 0.1184 & -0.2570 & -0.2176& -0.2220 & 0.0134 \\
(6,2) & 0.2891 & 0.1421 & -0.3034 & 0.1613 & -0.2621 & -0.2180 & -0.2214 & 0.0133 \\
(5,3) & 0.2799 & 0.1908 & -0.2694 & 0.0784 & -0.2609 & -0.2271 & -0.2158 & 0.0135 \\
(6,1) & 0.2651 & 0.0697 & -0.2543 & 0.1845 & -0.2703 & -0.2181 & -0.2259 & 0.0131 \\
(5,2) & 0.2521 & 0.2365 & -0.4419 & 0.2048 & -0.2777 & -0.2206 & -0.2235 & 0.0134 \\
(4,3) & 0.2447 & 0.1916 & -0.2325 & 0.0406 & -0.2697 & -0.2410 & -0.2142 & 0.0124 \\
(5,1) & 0.2279 & 0.1396 & -0.4372 & 0.2968 & -0.2929 & -0.2199 & 0.2338 & 0.0134\\
(4,2) & 0.2159 & 0.3783 & -0.6042 & 0.2239 & -0.3022 & -0.2300 & -0.2258 & 0.0129 \\
\end{tabular}
\end{center}
\end{table}

{\color{green}
\begin{table}[h!]
\begin{center}
\caption{Roll-up energy and dihedral contribution to energy for chiral CNTs.}
\label{tab:chiralen}
\vskip 3pt
\begin{tabular}{ccc}
 \\[3pt]
 &  Roll-up energy & Dihedral contribution \\[3pt]
$(n,m)$ &  (ev/atom) &( \%) \\[3pt]
\hline
(20,1) &  0.0280 & 49.7\\
(11,10) &  0.0355 & 49.7 \\
(10,1) &  0.1029 & 47.0\\
(6,5) & 0.1240 & 46.7\\
(6,3) &  0.1762 & 44.7 \\
(7,1) & 0.1941 & 43.8\\
(6,2) & 0.2115 &  43.3\\
(5,3) & 0.2224 & 43.2 \\
(6,1) &  0.2529 & 41.7\\
(5,2) &  0.2765 & 41.1\\
(4,3) & 0.2872 & 41.0\\
(5,1) & 0.3435 & 38.6\\
(4,2) &  0.3746 & 37.8\\
\end{tabular}
\end{center}
\end{table}
}

\cleardoublepage

\end{document}